\documentclass[11pt,a4paper,twoside,dvips]{article}

\usepackage[a4paper,hmargin=25mm,vmargin=25mm]{geometry}%

\usepackage{graphicx}
\usepackage{color}

\usepackage{pstricks}
\usepackage{psfrag}

\usepackage{tikz}
\usetikzlibrary{calc,intersections,through,backgrounds}

\usepackage[nottoc]{tocbibind}
    
\usepackage[nodayofweek]{datetime} 
\newdateformat{mydate}{ \monthname[\THEMONTH], \THEYEAR}

\usepackage[colorlinks,breaklinks=true]{hyperref}

\hypersetup{
colorlinks,citecolor=violet,linkcolor=violet,urlcolor=blue
}

\usepackage{breakurl}

\usepackage{scrextend}
\deffootnotemark{\textsuperscript{*\thefootnotemark}}
\deffootnote{2em}{1.6em}{*\thefootnotemark\enskip}

\usepackage[reqno]{amsmath}
\usepackage{amsthm}
\usepackage{amsfonts}
\usepackage{amssymb}
\usepackage{bm}
\numberwithin{equation}{section}

\newcommand{\rmd}{\text{d}}
\newcommand{\del}{\partial}

\newcommand{\order}[1]{\mathcal{O} \left( #1 \right)}

\newcommand{\Rbb}{\mathbb{R}}
\newcommand{\Sbb}{\mathbb{S}}
\newcommand{\Tbb}{\mathbb{T}}
\newcommand{\Bmbb}{\mathbb{B}}
\newcommand{\Zbb}{\mathbb{Z}}
\newcommand{\Abb}{\mathbb{A}}

\theoremstyle{theorem}

\theoremstyle{remark}

\theoremstyle{definition}

\newcommand{\ie}{i.e.}

\newcommand{\PO}{periodic orbit }
\newcommand{\TS}{transition state }
\newcommand{\DS}{dividing surface }
\newcommand{\TM}{transition manifold }
\newcommand{\DM}{dividing manifold }

\newcommand{\TSs}{transition states }
\newcommand{\DSs}{dividing surfaces }

\newcommand{\DoF}{degree of freedom }
\newcommand{\DoFs}{degrees of freedom }

\begin{document}


\begin{center} 
{\huge Morse bifurcations of transition states in bimolecular reactions}\\[10pt]

{\large R S MacKay$^{1 \; 2}$ and D C Strub$^{1}$}\\[10pt]

$^1$ Mathematics Institute and $^2$ Centre for Complexity Science, University of Warwick, Coventry CV4 7AL, U.K.\\[20pt]


\end{center}

\begin{abstract}

The transition states and dividing surfaces used to find rate constants for bimolecular reactions are shown to undergo Morse bifurcations, in which they change diffeomorphism class, and to exist for a large range of energies, not just immediately above the critical energy for first connection between reactants and products.
Specifically, we consider capture between two molecules and the associated transition states for the case of non-zero angular momentum and general attitudes.
The capture between an atom and a diatom, and then a general molecule are presented, providing concrete examples of Morse bifurcations of transition states and dividing surfaces.

\end{abstract}

\\[20pt]

\noindent
{\small Published in
{\bf Nonlinearity}, \href{http://dx.doi.org/10.1088/0951-7715/28/12/4303}{28(12):4303-4329}}
\\[20pt]



\tableofcontents




\newpage

\section{Introduction}

One way of finding rates of reaction is to consider rates of transport in a low dimensional Hamiltonian system representing the specific reaction. 
Some of the first examples studied using \TS theory consisted of 
bimolecular reaction in gaseous phase, i.e.
\[A + B \rightarrow \text{products},\] 
for two (polyatomic) molecules $A$ and $B$.
Provided
the Born-Oppenheimer approximation holds, we can pass from the quantum mechanical system to a classical one, namely the Hamiltonian system for the motion of the nuclei interacting via a potential given by the (ground state) energy of the electrons\footnote{Assumed non-degenerate and hence a smooth energy function, without conical singularities, see e.g. \cite{Domcke2004}.} as a function of the internuclear coordinates.
Then, if this extremely high dimensional Hamiltonian system is, at any instant, the product of ``reacting'' two molecule sub-systems that are independent of each other, we may consider the evolution of an ensemble of individual reactions in this low dimensional Hamiltonian sub-system. 
For this, we require the gas to be sufficiently dilute. See Keck \cite{Keck1967} for a nice review of \TS theory and discussion of the assumptions involved.
Finally, since the low dimensional system is Hamiltonian and energy is conserved, we can restrict our attention to the energy levels, and consider the flow of ergode\footnote{\emph{Ergode} is Boltzmann's name for a microcanonical ensemble, see \cite[pages 242, 367]{Brush}. We shall
use it interchangeably with energy-surface volume.}, as a function of the energy.
Finding (microcanonical) reaction rates translates to finding the rate of transport of ergode between regions representing reactants and products. 

Transition state theory provides upper bounds on rates of transport via the flux of ergode through a \emph{dividing surface} that separates the regions of interest. The \DS approach, which we reviewed in \cite{MacKay2014}, can be traced back to the works of Marcelin \cite[Chapter 2]{Marcelin} and Wigner \cite{wigner1937calculation}. 
In order to obtain a useful upper bound, the dividing surface is chosen to have locally minimum flux in a given direction. For this to be the case, the dividing surface must be the union of surfaces of unidirectional stationary flux with no local recrossings, which in turn is the case for surfaces spanning a closed, invariant, orientable, codimension-2 submanifold of the energy level, known as a \emph{transition state}.
The simplest transition states and dividing surfaces are found in the basic transport scenario of ``flux over a saddle'' \cite{MacKay1990}.  
These are transition states diffeomorphic to $\Sbb^{2m-3}$ for energies just above the index-1 critical point of the Hamiltonian for $m$ degree of freedom systems, with dividing surfaces diffeomorphic to $\Sbb^{2m-2}$. 
For two \DoF systems, these \TSs are hyperbolic periodic orbits. 
As the energy is increased, they may lose normal hyperbolicity and bifurcate. 
The bifurcations of periodic orbits are well known, see e.g.~\cite[Section 8.6]{abraham1987foundations}, and those of periodic orbit transition states have been studied for a long time, see e.g. \cite{DeVogelaere1955, Pollak1978, Davis1987}.
Instead, for higher \DoFs than two the \TSs for the basic scenario are normally hyperbolic $2m-3$ spheres (within the energy levels) and very little is known about their bifurcations.

What had been overlooked, and was explained in \cite{MacKay2014}, is that the \TSs may lose normal hyperbolicity by becoming singular, i.e.~the manifold structure fails, at a higher critical energy value but then regain their normal hyperbolicity for values above the critical energy.
These are \emph{Morse bifurcations} \cite[Appendix B]{MacKay2014}, leading to a change in diffeomorphism class of the transition states and dividing surfaces spanning them.
They occur when the union of \TSs over a range of energies forms a smooth normally hyperbolic submanifold of state space, which we call a \emph{transition manifold}, and there is a critical value of the Hamiltonian function restricted to the transition manifold. By definition, the \TSs are the level sets of this restricted Hamiltonian and undergo a bifurcation. Similarly, the dividing surfaces, which span the transition manifold, undergo a Morse bifurcation.
Actually, they occur because the energy levels undergo a Morse bifurcation themselves so the dividing surfaces, and therefore the transition states, must also undergo a change of diffeomorphism class in order to still separate these in two.
Morse theory tells us that we can continue the \TSs and dividing surfaces, with respect to the energy, through critical values, as well as giving the diffeomorphism class of the submanifolds. This is useful, for example, when the \TSs above some critical energy are disjoint, which we shall see in the atom-frozen diatom system, as we then know exactly what objects to look for.
Recently, there has been a lot of interest in general transport problems (as opposed to the basic one), non-minimum energy paths \cite{Osborn2008}, and roaming reactions \cite{Bowman2011}.
These transport scenarios may emerge from the basic one via a Morse bifurcation, in which case by
considering these bifurcations, we obtain \TSs and \DSs that can be used to find the rates of transport.

We shall consider transport between the region representing two distant molecules (reactants) and the region in which the molecules are close. The latter does not however generally constitute the products\footnote{Association and recombination reactions are largely limited to reactions in condensed phase or solvent, see e.g.~\cite[Chapter 1]{Henriksen2008}}. This is the \emph{capture} transport problem associated with the necessary first step of the molecules getting close enough to react.
The \emph{capture rate} (sometimes also called \emph{collision rate}) provides an upper bound on the reaction rate,
as we do not expect all captured trajectories to proceed to the products region \cite{Chesnavich1980}.
Note that there might actually be multiple product regions, but for a two-body capture process there is only one final region of interest.

Capture rates are crucial for many physical processes, and have a long history dating back at least to 1905 with Langevin's early contribution \cite{Langevin1905}, see review by Chesnavich and Bowers \cite{Chesnavich1982}.
Two assumptions are usually found in the literature.
Firstly, the reacting Hamiltonian systems are assumed to have Euclidean symmetry, that is to be invariant under translations and rotations. This is the case for gas phase reactions with no background (electro-magnetic) field.
The Hamiltonian system can then be reduced to a family of systems, in centre of mass frame, parametrised by the angular momentum.
The Hamiltonian function then contains both Coriolis and centrifugal terms.
Secondly, the energy is taken to be below those at which the two molecules dissociate and centrifugal and Coriolis forces to be sufficiently weak such that the molecules are well defined and in the small vibrations regime\footnote{We are implicitly assuming that the molecules are \emph{normal}, i.e.~that their potential has non-degenerate critical points corresponding to rigid equilibrium configurations. Molecules that are not normal are \emph{anomalous}. We avoid the term \emph{rigid}, as it might lead to confusion with the rigid body limit with no vibrations.}. 
These assumptions allow to distinguish between intermolecular degrees of freedom (distance and relative attitudes of the molecules) and intramolecular ones.
We too shall consider systems that satisfy these assumptions.

We want to find the rate of capture, which we shall assume can be thought of as transport between regions on either side of a non-degenerate maximum $\bar x_c$ of the effective potential (centrifugal terms plus potential) with respect to the intermolecular distance $x$. 
In the literature, this maximum is generally a centrifugal maximum obtained by balancing the repulsive centrifugal terms with the attractive long distance potential energy.
Alternatively, $\bar x_c$ could be a non-degenerate chemical maximum of the bimolecular potential and therefore of the effective potential for small angular momentum. 
Provided $\bar x_c$ is sufficiently large, such that capture occurs in a region where the potential is only weakly dependent upon the attitudes of the molecules, and sufficiently non-degenerate, we shall see that fixing the intermolecular distance degree of freedom to the maximum value gives a normally hyperbolic transition manifold in state space, which can be spanned by a dividing manifold.
The restriction of these manifolds to the energy levels gives dividing surfaces and transition states, which we shall refer to as \emph{capture} transition states. The literature often refers to them as \emph{orbiting} or \emph{loose} transition states \cite{Chesnavich1982,Pechukas1976}.

The central field model, in which the attitudes of the colliding pair are ignored, is attributed to Langevin \cite{Langevin1905}.
In this very early work, one already finds capture periodic orbit transition states. Langevin considers the capture process using scattering theory. 
For an introduction to scattering theory see e.g. \cite[Section 3.10]{Goldstein2002}, whereas for a comparison with the \DS approach see \cite{Chesnavich1982}.
Non-central fields were considered later, also usually from a scattering theory perspective, starting with the works of Pechukas \cite{Pechukas1980} and Chesnavich, Su and Bowers \cite{Chesnavich1980}.
As the energy is increased above the critical energy for first connection between reactants and ``products'', the intermolecular attitude and angular momentum degrees of freedom will generally be involved in interesting sequences of Morse bifurcations
causing changes in the diffeomorphism class of the energy levels, as the ``cone of acceptance'' (i.e.~orientations for which capture is possible) opens up,
and therefore changes of the capture transition states and dividing surfaces.
Instead, the intramolecular degrees of freedom are assumed to consist of small vibrations, thus not playing a role in the Morse bifurcations. 
Using Morse theory, one can know which the correct \TSs and dividing surfaces are, as we shall outline for two examples, and then study rates of transport for larger energies than previously thought possible.
Here, we shall focus on the Morse bifurcations, the reduction and the details of bimolecular capture problems.
Some analysis of ``reaction dynamics'' of rotating molecules has been done recently in \cite{Ciftci2012, Kawai2011a}, but we are interested in the interaction of two rotating molecules.

Whether captured pairs then go on to react can be thought of as a further transport problem with its own (\emph{reaction})
transition states and dividing surfaces, possibly (but not necessarily) associated to other maxima $\bar x_i$ of the effective potential. These are often referred to as \emph{tight} transition states.
The capture and reaction transition states are therefore in series. The simplest case will be when these are distant and the level sets of separate transition manifolds. However, 
even when ``separate'' their stable and unstable manifolds, which act as transport barriers, will intersect, determining the ``reaction channels''.
Some trajectories joining reactants and products might roam in the region between the (capture and reaction) dividing surfaces, that is follow trajectories with a non-monotonic intermolecular distance in time, before finally crossing the reaction dividing surface. This is to be expected due to coupling between degrees of freedom, and was recently given as an explanation of what has been called \emph{roaming reactions} \cite{Mauguiere2013}.

Reaction rates have an equally long history as capture rates, and bimolecular reactions have played the role of test problems since the early days of transition state theory (as noted in \cite{Wigner1938}). 
The transport problems associated with reaction tend to be harder because the chemical potentials are at best not simple and often degenerate. 
Similarly to how the first capture models were simplified by making the fields central, reaction rates were first, and largely still, considered for collinear and planar systems with zero angular momentum.
Note that reaction transition states will also generally undergo Morse bifurcations.

Hamiltonian systems with symmetry, such as the molecular $n$ body systems with Euclidean symmetry considered here, have conserved quantities by Noether's theorem (see e.g.\ \cite[Appendix 5]{arnold1989mathematical}, \cite[Section 2.7]{Marsden1992}), and can be reduced via symplectic reduction.
The Euclidean symmetry group of our $n$ body systems is a semi-direct product $SE(3) = \Rbb^3 \ltimes SO(3)$ and can be reduced in stages \cite[Chapter 4]{Marsden2007}.
The translational symmetry and associated linear momentum are easily reduced, whereas the rotational symmetry, which does not act freely on the whole of state space, and the associated angular momentum require singular reduction, see e.g. \cite{Sjamaar1991}, \cite[Section 8]{Ortega2004}.
The reduced state space is therefore a stratified symplectic manifold, however the Hamiltonian flow leaves the connected components of the strata invariant so the reduced dynamics can be studied on each stratum individually. 
The stratified symplectic manifold for our reduced molecular $n$ body systems is composed of three strata. Both singular strata have zero angular momentum, whereas on the principal stratum the angular momentum is arbitrary. This is discussed in \cite{Strub2015}.
Actually, reactions with initial positions and momenta confined to a plane remain in this plane for all successive times, and 
reduced planar systems have no angular momentum degree of freedom and no coordinate singularities at collinear configurations. 
In Section \ref{planarAtomDiatom}, we therefore consider planar capture between an atom and a diatom and find our first examples of Morse bifurcations. This example and that of two interacting diatoms were considered in less detail in \cite{MacKay2014}.
A spatial example, of an atom interacting with a molecule, is considered in Section \ref{spatialBimol}.
We choose to use the coordinates that one obtains via the gauge theoretic approach to cotangent bundle reduction as outlined by Littlejohn and Reinsch \cite{Littlejohn1997}, and reviewed in Appendix \ref{biGauge}, where we
discuss the symplectic form and consider the transformation to Serret-Andoyer coordinates on the angular momentum sphere, which is not done in \cite{Littlejohn1997}.
This approach gives coordinates, which are furthermore physically meaningful since coordinates and momenta are not mixed, however it involves considering the rotational $SO(3)$ action on configuration space as opposed to the lifted action in state space, and therefore introduces coordinate singularities. These are avoided by considering non-collinear molecules. Ways to deal with them are discussed in the Conclusions in Section \ref{conclude}.


\section{Planar atom-diatom reactions}
\label{planarAtomDiatom}

\begin{figure}
\begin{center}
	\def\JPicScale{0.5}
	\input{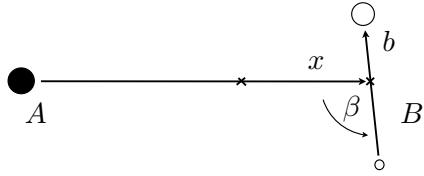}
\end{center} 
\caption{Choice of Jacobi vectors and reduced coordinates $q = (x,b,\beta)$ for planar atom-diatom reactions.}
\label{jacobiFig}
\end{figure}

The simplest non-trivial capture example is planar reactions between an atom and a diatom. 
The planar reduced three-body Hamiltonian system that represents an atom interacting with a diatom with no background field is the family of mechanical systems\footnote{A \emph{simple mechanical system} is a Hamiltonian system whose state space is the cotangent bundle of a Riemannian manifold (\emph{configuration space}) with canonical symplectic form, and the Hamiltonian is the sum of the positive definite \emph{kinetic energy}, given by the metric, and a \emph{potential energy}.}
$(T^*\Rbb^3_+, \omega, H)$, parametrised by the angular momentum $\lambda \in \Rbb$, with
\begin{align*}
H(z; \lambda) &= 
\frac{1}{2} \left( \frac{1}{m} p_x^2 + \frac{1}{m_b} p_b^2 + \left( \frac{1}{m_b b^2} + \frac{1}{m x^2} \right) \left( p_\beta - \frac{m_b b^2}{m x^2 + m_b b^2} \lambda \right)^2 \right)
+ V (q; \lambda), \\
V(q; \lambda) &= 
\frac{1}{2} \frac{\lambda^2}{m x^2 + m_b b^2}
+ U(q)
\end{align*}
where $z = (q,p)$, $q = (x,b,\beta)$,
$x$ is the intermolecular distance, $\beta$ the attitude and $b$ the length of the diatom, see Figure \ref{jacobiFig}. $V$ is the effective potential with the centrifugal term. 
The parameters are the reduced masses $m$ and $m_b$, and the magnitude of the angular momentum $\lambda$. 

We shall restrict our attention to energies below that at which the diatom dissociates, so we have a two-body capture problem. 
Specifically, we shall consider the capture of an atom by a strongly bonded diatom that is distant and rotating slowly, assuming that the potential of the pair has a non-degenerate maximum $\bar x_c$ with respect to $x$ for large $x$, and that the centrifugal and Coriolis forces are not too large. 
Thus, for energies below that at which the diatom dissociates, the reduced coordinates and their momenta split into intermolecular $(x,\beta)$ and intramolecular $b$ degrees of freedom.
Our assumptions imply that the diatom is in the small vibrations regime, i.e.\ $(b,p_b)$ is an elliptic degree of freedom, and rotating slowly, so in the neighbourhood of $\bar x_c$ we will find a codimension-2  normally hyperbolic submanifolds $N$ and span it by a dividing manifold $S$. This will be the capture \DM that trajectories must cross in order for capture to occur.
The assumptions will be introduced by scaling the coordinates, after which the dynamics and the existence of the transition manifold $N$ will be clear.
Then, we will show that the level sets of $N$ and $S$, the transition states $N_E$ and \DSs $S_E$, generically undergo Morse bifurcations involving the $(\beta, p_\beta)$ degree of freedom, as the energy is varied.

For simplicity, the masses of the atom and the diatom are set to $m = m_b = 1$.
A better approach would be to non-dimensionalise the variables. 
Note that normalised or mass-weighted Jacobi coordinates do not remove the mass dependence.

\begin{figure}
\begin{center}
	\def\JPicScale{0.4}
	\input{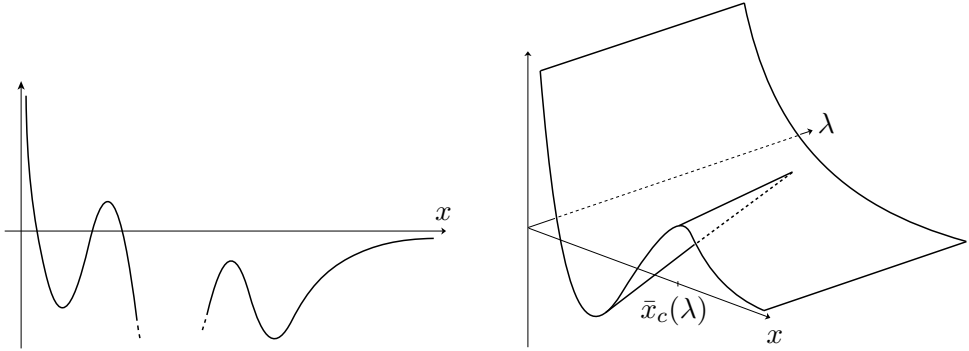}
	\end{center}
	\caption{Left: Typical graph of molecular potential restricted to the intermolecular distance, with repulsive short range, attractive long range and extrema in between. Right: Example graph of the effective potential over the intermolecular distance and angular momentum $(x,\lambda)$ showing the disappearance of the intermolecular maximum $\bar x_c$, via centre-saddle bifurcations.}
	\label{atomdiatomXPot}
\end{figure}

The intermolecular terms of the potential will be repulsive at short ranges, possibly have a number of maxima (and therefore minima) with respect to the intermolecular distance $x$ in the mid ranges, and be attractive at long ranges with inverse $k$-power of $x$ terms \cite{Stone2013}, see Figure \ref{atomdiatomXPot}.
The potential is then summed to the repulsive centrifugal term to give the effective potential.
In the long (physical) range,
provided the attractive potential falls off faster than the centrifugal potential as $x \rightarrow \infty$, i.e.~$k>2$, 
the effective potential has a centrifugal maximum $\bar{x}_\lambda (b, \beta; \lambda)$. 
In the short (chemical) range of the potential, the chemical maxima of $U$ with respect to $x$ imply chemical maxima of $V$ for $\lambda$ small with respect to the slope of $U$ at the maxima.
In either case, as $\lambda$ increases the maxima
will ``collide'' with the minima and disappear, see Figure \ref{atomdiatomXPot}.

We shall think of capture as transport between the regions on either side of the largest maximum $\bar x_c (b, \beta; \lambda)$, assumed large with respect to the length of the diatom $b$.
We introduce the capture scale by setting
$b = \varepsilon_c \tilde b$ and $p_b = \varepsilon_c^{-1} \tilde{p_b}$ with $ 0 < \varepsilon_c \ll 1$, and taking $\bar x_c$ to be of order 1. Essentially, $\varepsilon_c$ is the ratio between the size of the diatom and its distance to the atom, however in practice we assume
that for distant pairs, the potential energy is weakly dependent on the attitude of the diatom, since the pair are distant, and we can choose $\varepsilon_c$ such that
\[ U(q; \varepsilon_c) =  \varepsilon_c^{-2} U_b(\tilde b) + U_c^0(x) + \varepsilon_c^{2} U_c^2(q; \varepsilon_c). \]
This is the case for potentials that are inverse power functions of the intermolecular distance, which can be expanded using Legendre polynomials.
The Hamiltonian function expands into
\begin{align*}
H(z; \lambda, \varepsilon_c) &= 
\varepsilon_c^{-2} \left( \frac{1}{2} p_b^2 + \frac{1}{2} \frac{p_\beta^2}{b^2}  + U_b(b) \right)
+ \frac{1}{2} p_x^2 + \frac{1}{2} \frac{\left( p_\beta - \lambda \right)^2}{ x^2} + U_c^0(x) \\
&+ \varepsilon_c^2 U_c^2(q; 0) + \order{\varepsilon_c^4},
\end{align*}
where we have dropped the tildes. We note that $\beta$ does not appear until order $\varepsilon_c^2$, so $p_\beta = \lambda_\beta + \order{\varepsilon^2_c}$ with constant $\lambda_\beta$.
Furthermore, the system separates into slow and fast degrees of freedom, i.e.~at order $\varepsilon_c^{-2}$ we find the fast oscillations of the diatom plus a ``centrifugal'' term, at order $\varepsilon_c^0$ there is the intermolecular (capture) dynamics, and then there are the higher order terms.
Up to order $\varepsilon_c^0$, the $x$ and $b$ degrees of freedom are uncoupled, and $p_\beta = \lambda_\beta$.

The diatom will have an equilibrium configuration, which corresponds to a non-degenerate minimum $\bar{b}(x,\beta)$ of the joint atom-diatom potential with respect to the intramolecular distance $b$. We assume this minimum to be highly non-degenerate, i.e.~that the diatom is strongly bonded, and that the centrifugal and Coriolis forces on the diatom are not too strong, so the diatom will vibrate about its equilibrium without significant distortion.
We therefore, linearise $b$ about $\bar b_0$ by setting $b = \bar b_0 + \varepsilon_b \tilde b$ and $p_b = \varepsilon_b^{-1} \tilde{p_b}$ with $ 0 < \varepsilon_b \ll 1$. The constant $\varepsilon_b$ is chosen such that
\[
U(q; \varepsilon) 
=  \varepsilon_c^{-2} \left( \bar U_b^0 + \frac{1}{2} \varepsilon_b^{-2} \bar U_b^2 \tilde b^2 \right) + U_c^0(x) + \varepsilon_c^{2} U_c^2(q; \varepsilon_c) + \order{\varepsilon_b^3},
\]
where $\bar U_b^0 := U_b (\bar b_0)$, $\bar U_b^2 := \varepsilon_b^{4} \del_{bb}^2 U_b (\bar b_0)$ is order one, and $\varepsilon = (\varepsilon_c, \varepsilon_b)$.
Recall that we are assuming $ \del_{bb}^2 U_b (\bar b_0)$ to be large.
This scaling ensures that the leading order terms of the potential with respect to the coordinates are of the same order as their conjugate momenta in the kinetic energy.
The Hamiltonian function, again dropping the tildes, becomes
\begin{align*}
H(z; \lambda, \varepsilon) &= 
\varepsilon_c^{-2} \varepsilon_b^{-2} \frac{1}{2} \left(  p_b^2 + \bar U_b^2 b^2  \right) 
+ \varepsilon_c^{-2} \frac{1}{2 \bar b_0^2} p_\beta^2
+ \left( \frac{1}{2} p_x^2 + \frac{1}{2 x^2} \left( p_\beta - \lambda \right)^2 + U_c^0(x) \right) \\
&+ \varepsilon_c^2 U_c^2(\bar b_0,x,\beta; 0) + \order{\varepsilon_c^4, \varepsilon_b^1}.
\end{align*}

We shall further simplify our Hamiltonian by setting $p_\beta = \varepsilon^2_c \tilde p_\beta$, i.e.~considering $\lambda_\beta = 0$. 
General $p_\beta$ is considered in the disconnecting example of \cite{MacKay2014}.
The scaled system consists of
\begin{align*}
H(z; \lambda, \varepsilon) &= 
\varepsilon_c^{-2} \varepsilon_b^{-2} \frac{1}{2} \left(  p_b^2 + \bar U_b^2 b^2  \right) 
+  \frac{1}{2} p_x^2 + \frac{1}{2 x^2} \lambda^2 + U_c^0(x)  \\
&+ \varepsilon_c^2 \left( \frac{1}{2 \bar b_0^2} p_\beta^2 + \frac{1}{x^2} p_\beta \lambda + U_c^2(\bar b_0,x,\beta; 0) \right) + \order{\varepsilon_c^4, \varepsilon_b^1}
\end{align*}
and
\[ \omega = \rmd b \wedge \rmd p_b + 
\varepsilon_c^2 \rmd \beta \wedge \rmd p_\beta + \rmd x \wedge \rmd p_x,  \]
from which we get the equations of motion up to order $\varepsilon^0$, namely
\begin{align*}
\dot b &= \varepsilon_c^{-2} \varepsilon_b^{-2} p_b, &
\dot \beta &= \frac{1}{\bar b_0^2} p_\beta - \frac{1}{x^2} \lambda , & 
\dot x &= p_x, \\
\dot p_b &= - \varepsilon_c^{-2} \varepsilon_b^{-2} b , &
\dot p_\beta &= - \del_\beta U_c^2 (\bar b_0, \beta, x; 0) , &
\dot p_x &= \frac{1}{x^3} \lambda^2 - \del_x U_c^0 (x).
\end{align*}

By assumption, the intermolecular distance degree of freedom is hyperbolic, and the intramolecular distance is in the small vibrations regime, i.e.~elliptic. These dynamics are uncoupled to this order. 
As the diatom rotates, the attitude degree of freedom will display both kinds of motion. 
Provided the $(x, p_x)$ degree of freedom is more strongly hyperbolic that the $(\beta, p_\beta)$ one, that is
the maximum $\bar x_c$ is sufficiently non-degenerate,
the submanifold
\[N_0 = \lbrace z \in M_\lambda \vert x = \bar x_c^0 (\lambda), \; p_x = 0 \rbrace \]
is almost invariant, or more precisely invariant to order $\varepsilon^0$, and normally hyperbolic.
By normal hyperbolicity theory, there is a true normally hyperbolic submanifold $N$ nearby. 
Given $N_0$, we could find a better approximation to $N$ as explained in \cite{MacKay2014}.
However, for the purpose of studying the Morse bifurcations of the transition states, $N_0$ is a sufficiently good approximation.

The normally hyperbolic submanifold $N$ is a transition manifold, as it can be spanned by a dividing manifold.
The approximate transition manifold $N_0$ is spanned by
\[S_0 = \lbrace z \in M_\lambda \vert x = \bar x_c (b, \beta; \lambda) \rbrace.\]
The transition states $N_E$ and dividing surfaces $S_E$ are then approximately the level sets of the Hamiltonian restricted to the approximate transition and dividing manifolds.
As the energy varies, we expect these to bifurcate. The transition states may lose normal hyperbolicity. For atom-diatom reactions, $N_E$ are 3-dimensional manifolds, so it is not well understood how they lose normal hyperbolicity.
Freezing the diatom, the system only has two \DoFs and $N_E$ is a periodic orbit. In \cite{MacKay2014}, we saw that these disappear in a centre-saddle bifurcation. However, before the loss of normal hyperbolicity, the capture \TSs will undergo
changes of diffeomorphism class via Morse bifurcations. 

If we write
\[N = \lbrace z \in M_\lambda \vert x = \bar x_c^0 (\lambda) + \varepsilon_c^2 \bar x_c^2 (z) + \order{\varepsilon_c^4}, \; p_x = 0 + \varepsilon_c^2 P_2 (z) + \order{\varepsilon_c^4} \rbrace, \]
we find that the Hamiltonian function restricted to the transition manifold $N$ is independent of $\bar x_c^2$ and $P_2$ up to order $\varepsilon_c^2$, namely
\[
H_N(z; \lambda, \varepsilon) = 
\varepsilon_c^{-2} \varepsilon_b^{-2} \frac{1}{2} \left(  p_b^2 + \bar U_b^2 b^2  \right) 
+ \varepsilon_c^2 \left( \frac{1}{2 \bar b_0^2} v_\beta^2 + U_c^2(\bar b_0,\bar x_0^c,\beta; 0) \right) + \order{\varepsilon_c^4, \varepsilon_b^1},
\]
where $v_\beta = p_\beta - \frac{\bar b_0^2}{\bar x_0^2} \lambda + \cdots$ is the non-canonical momentum, and we dropped constant terms.

\begin{figure}
\begin{center}
\def\JPicScale{0.5}
	\input{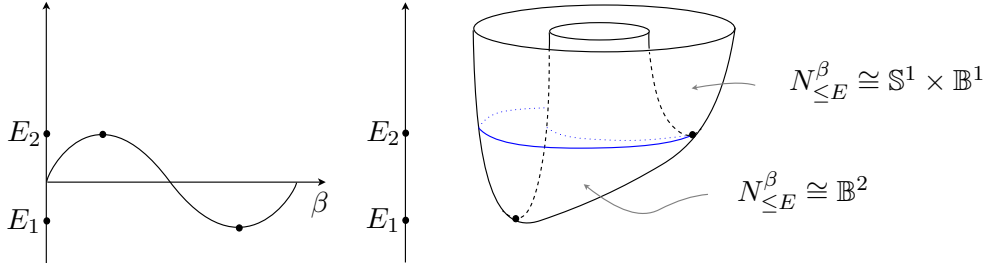}
\end{center}
\caption{Graphs of an example frozen Hamiltonian restricted to the transition manifold $H^{\beta}_{N}$ (right) and the potential $U_c^2$ (left) for an atom-diatom reactions with the atom attracted to one of the sides of the diatom (e.g. ion plus dipole), with intermolecular and diatom distances fixed.}
\label{atomDiatom}
\end{figure}

For $E$ below that at which the diatom dissociates, the intramolecular degree of freedom contributes only positive-definite terms to the restricted Hamiltonian function and
is not involved in any Morse bifurcations, which can be studied by considering the simpler function obtained by freezing the diatom, i.e.~minimising $H_{N}$ over $(b,p_b)$ by setting $b = \bar b_0 + \text{h.o.t.}$ and $p_b = 0$ giving
\[
H_N^\beta(\beta , p_\beta; \lambda, \varepsilon) = 
\varepsilon_c^2 \left( 
\frac{1}{2 \bar b_0^2} v_\beta^2 +  U_c^2(\bar b_0, \beta, \bar x_0; 0) \right) + \order{\varepsilon_c^4, \varepsilon_b^1}.
\]

Different reactions, i.e.~choices of atom and diatom, will have different potentials and different sequences of Morse bifurcations. An example frozen restricted Hamiltonian $H^{\beta}_{N}$ is depicted in Figure \ref{atomDiatom}.
Considering this $H_N^\beta$, we find that $N^{\beta}_{\leq E} = (H_N^\beta)^{-1} ((-\infty , E])$ bifurcates from $\Bmbb^2$ to $\Sbb^1 \times \Bmbb^1$, and so the level sets $N^{\beta}_{E}$ from $\Sbb^1$ to $\Sbb^0 \times \Sbb^1$, where $\Bmbb$ denotes a ball and $\Sbb$ a sphere.
Passing to the full system, we find that the transition manifold $N_{\leq E}$ bifurcates from $\Bmbb^4$ to $\Sbb^1 \times \Bmbb^3$, 
the transition state $N_E$ from $\mathbb{S}^3$ to $\mathbb{S}^1 \times \mathbb{S}^2$, and the dividing surface $S_E$ from $\mathbb{S}^4 $ to $ \mathbb{S}^1 \times \mathbb{S}^3$.
Thus, the energy levels are separated by the dividing surfaces for all energies, even above the critical one of the Morse bifurcation.
Other examples of $U_c^2$ are considered in \cite{MacKay2014}.

\begin{figure}
\begin{center}
\begin{pspicture}(11,4.5)
		\psfrag{a}{\tiny $E_1$}
		\psfrag{b}{\tiny $E_2$}
		\psfrag{f}[b][t]{\footnotesize $\phi_E$}
		\psfrag{E}[t][b]{\footnotesize $E$}
		\rput(2,2.1){\includegraphics[width=0.45\textwidth]{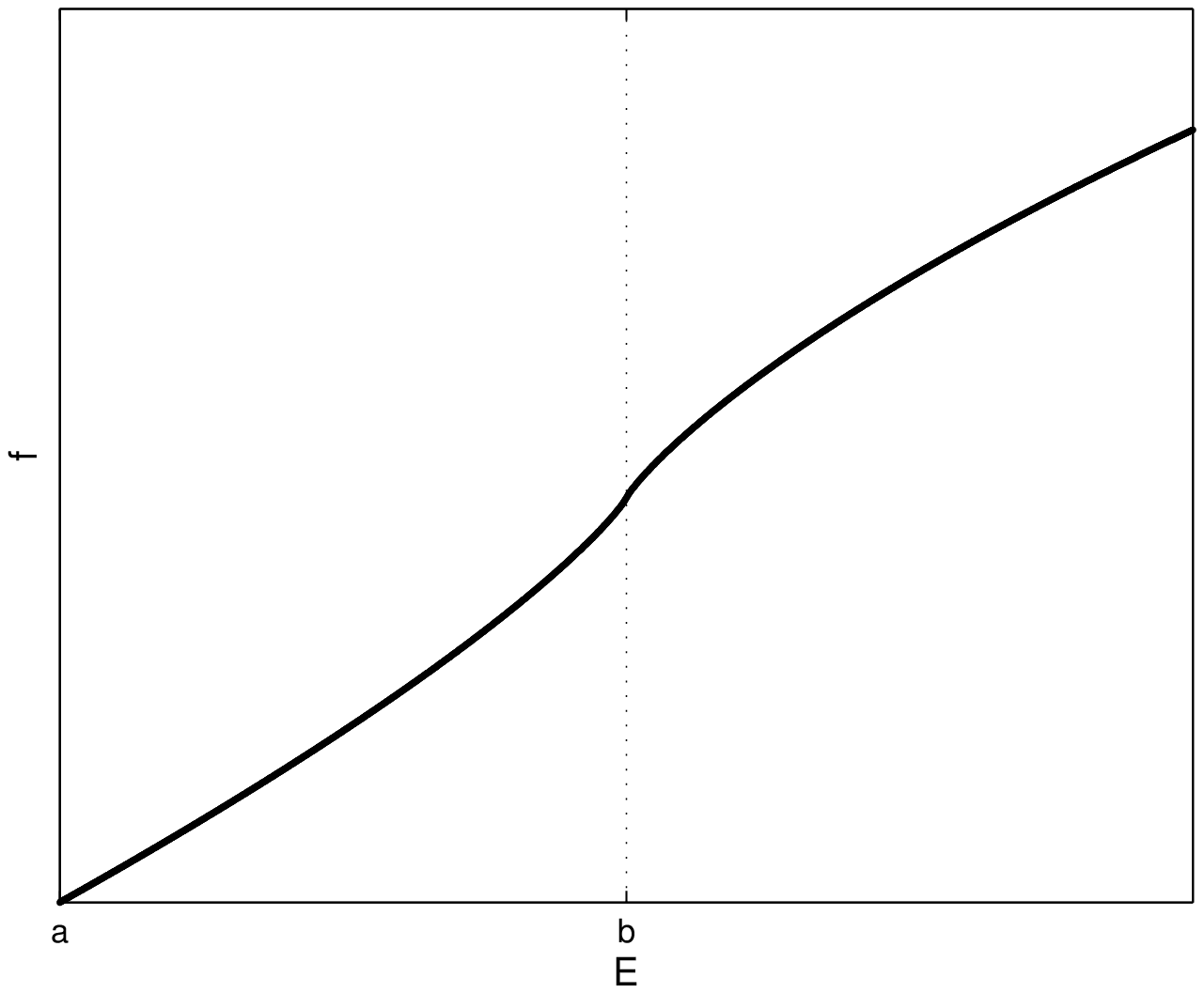}}
		\rput(8.7,2.1){\includegraphics[width=0.45\textwidth]{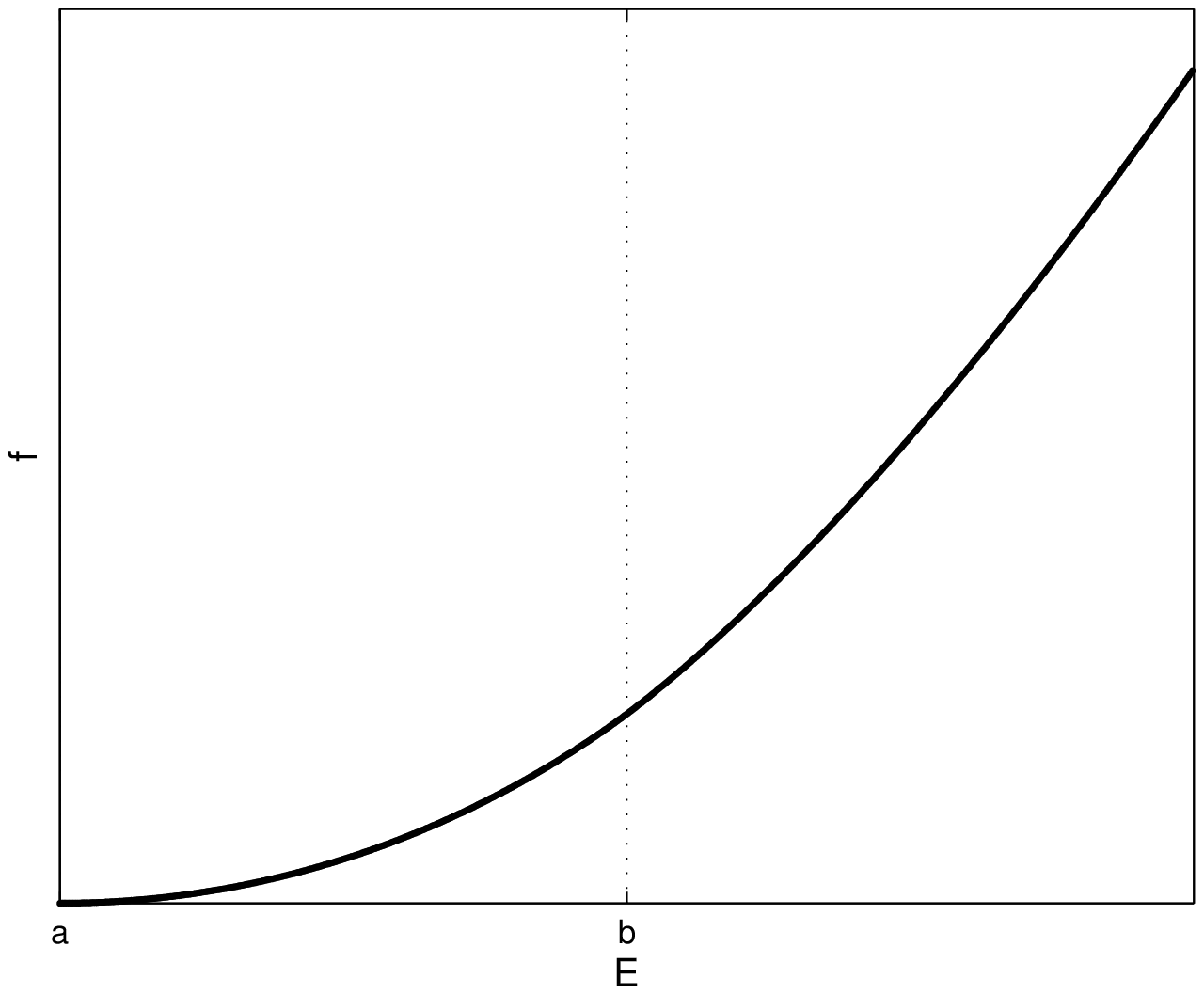}}
	\end{pspicture}
\end{center}
\caption{Example graphs of the flux $\phi_E$ as a function of energy $E$, for the capture transition state of the atom-diatom system, with potential as in Figure \ref{atomDiatom}, undergoing a disconnecting Morse bifurcation at $E_2$. Left: two \DoF system with frozen diatom. Right: three degree of freedom system.}
\label{fluxFig}
\end{figure}

Given a dividing surface $S_E$, we can compute the flux of ergode through it, denoted $\phi_E (S_E)$, and use this to find the rate constant.
The flux varies $C^{m-2}$ smoothly through the Morse bifurcations, for $m$ degrees of freedom, but not $C^{m-1}$ \cite{MacKay2014}. 
Graphs of $\phi_E$ as a function of $E$ for the capture dividing surface of the atom-diatom system with the example potential presented in Figure \ref{atomDiatom} can be seen in Figure~\ref{fluxFig}. 
For the \PO \TS of the two \DoF system with the frozen diatom, we see the log-like infinite slope singularity at the index-1 Morse bifurcation. This is not present in the graph of the flux through the 	\DS of the three-\DoF system.


\section{Spatial atom-molecule reactions}
\label{spatialBimol}

We now consider the capture of an atom by a 
polyatomic molecule.
The molecule shall be assumed to have a  non-degenerate non-collinear equilibrium, of its $n_b$ atoms, about which it is vibrating fast, and the system to have energy below that at which the molecule dissociates, so we can use the charts provided by the bundle approach to cotangent bundle reduction, see Appendix \ref{biGauge}.
The capture scenario will be the same as the atom-diatom one, modulo differences due to its spatial nature, such as an additional angular momentum degree of freedom.
We shall again assume that the effective potential has a non-degenerate maximum $\bar{x}_c$ with respect to the intermolecular distance $x$ when the pair is distant, that the attitude terms are not dominant in this range, and that $B$ is a normal molecule with a very non-degenerate equilibrium, i.e.\ that it is strongly bounded. In order to have $B$ in the (elliptic) small vibrations regime, and the attitude and angular momentum degrees of freedom dominated by the hyperbolic (in the neighbourhood of $\bar x_c$) intermolecular one, we ask that the molecule is rotating slowly with respect to the atom, that it is 3D with distinct principal moments of inertia and that the energy is bounded such that, roughly speaking, the angular momentum $l$ is almost perpendicular to the line of centres. 
These extra constraints due to the spatial nature shall be discussed in detail after we scale the intermolecular and intramolecular distance. Finally after also scaling the angular momentum coordinates and the attitude momenta, we shall find a normally hyperbolic transition manifold $N$ and \DM $S$ about $\bar x_c$ and discuss the possible Morse bifurcations of the \TSs $N_E$ and dividing surfaces $S_E$. These will now also involve the angular momentum degree of freedom, degeneracies when the moment of inertia tensor has equal moments, and require an understanding of the diffeomorphism class of the reduced space, which we shall discuss.

Capture between two non-collinear molecules can be studied following the same steps as those below, and will be presented in a future publication.

We denote the reduced molecular $n = n_b + 1$ body Hamiltonian system by $( \tilde M_\lambda , \omega_\lambda , H_\lambda)$. The state space $\tilde M_\lambda$ is the subset of the principal stratum with non-collinear configurations. It is a smooth $(6n_b-4)$-manifold diffeomorphic to a (generally non-trivial) $\Sbb^2$ fibre-bundle 
with base space $T^* ( Q_{I_d}/SO(3) )$, where $Q_{I_d}$ is the non-collinear subset of the translation reduced configuration space $Q \cong \Rbb^{3(n-1)}$, and fibre the angular momentum sphere $\Sbb^2_\lambda$ \cite[Section 2.3]{Marsden2007}.

In canonical coordinates $z = (q,q_\lambda, p, p_\lambda) \in \tilde M_\lambda$, we have
\begin{align*}
H \left( z; \lambda \right) &= \frac{1}{2} \sum_{i,j=1}^{3n-6} \sum_{k=1}^{3} ( p_i - A_{ik} (q) l_k (z_\lambda; \lambda) ) K^{ij} (q) ( p_j - A_{jk} (q) l_k (z_\lambda; \lambda) ) + V( q, z_\lambda; \lambda ) \\
V( q, z_\lambda; \lambda ) &= \frac{1}{2} \sum_{i,j=1}^{3} l_i(z_\lambda; \lambda) I^{ij} (q) l_j (z_\lambda; \lambda) + U \left( q \right). 
\end{align*}
where $q$ are the internal coordinates, which we shall choose shortly, $\lambda$ the magnitude of the angular momentum and $z_\lambda = (q_\lambda, p_\lambda)$ canonical Serret-Andoyer coordinates on the angular momentum sphere, such that 
e.g.~the angular momentum is given by
\[l (z_\lambda; \lambda)
 = ( p_\lambda, \sqrt{ \lambda^2 -  p_\lambda^2} \sin q_\lambda, \sqrt{ \lambda^2 - p_\lambda^2} \cos q_\lambda).\]
More than one chart is necessary, due to the coordinate singularities both for $Q_{I_d} /SO(3)$, with $n \geq 4$ bodies, as well as on the angular momentum sphere.
$V$ is the effective potential with the centrifugal terms, $K$ is the reduced metric, $I$ is the moment of inertia tensor, and $A$ is the gauge potential, present in the Coriolis terms. These are introduced in Apprendix \ref{biGauge}. 
We are uninterested in scaling effects due to the mass, so we set $m_i = 1$. This affects $K$ and $I$.

The rotating frame for the reduction is chosen such that the Jacobi vector along the line of centres $r_{n_b} (q)$ is parallel to the $x_1$ axis, and the remaining $SO(2)$ symmetry about the $x_1$-axis is used to orient the equilibrium configuration of the molecule $B$
such that its moment of inertia tensor $I_b^0 = \text{Diag} ( \mu_{b1}, \mu_{b2}, \mu_{b3} )$ with $\mu_{b1} > \mu_{b2} > \mu_{b3}$.
This determines the reduction of the $SO(2)$ symmetry about $x_1$ and the definition of the attitude of the molecule.
The most natural choice of reduced coordinates $q$ is the distance between atom and molecule $x$ and two angles $\beta  = (\beta_1, \beta_2) \in SO(3)/SO(2) \cong \Sbb^2$ for the attitude of the molecule, which are intermolecular coordinates, as well as $3(n_b - 2)$ coordinates $b$ for the intramolecular degrees of freedom of $B$, so $q = (x, \beta, b)$, unless $B$ has further symmetries of its own that can be reduced.
The intramolecular coordinates $b$ shall be chosen in order to simplify the Hamiltonian along the lines of the Eckart \cite{Eckart1935} and Sayvetz \cite{Sayvetz1939} conventions for normal and anomalous molecules in the small vibration regime. 
Essentially, we shall consider an Eckart convention for a normal molecule in the small vibrations regime interacting with an atom, for which the intermolecular coordinates are similar to the large amplitude coordinates of anomalous molecules considered by Sayvetz (see Apprendix \ref{spatialDetials}).
In the Eckart convention the rotations and vibrations are decoupled to leading order since the intramolecular coordinates $b$ are chosen to be Riemann normal coordinates for which the gauge potential $A_b(q)$ vanishes at the equilibrium configuration. This is discussed from a geometric perspective by  Littlejohn and Mitchell \cite{Littlejohn2002}. 

First, we scale the distances and their momenta.
We are interested in large $x$ about the (capture) maximum $\bar x_c$, so we scale $x  = \varepsilon_c^{-1} \tilde x$ and $p_x = \varepsilon_c \tilde  p_x$. 
Then, by passing to the intermolecular time, we scale the Hamiltonian such that $x$ is of order one.
Also, we consider the molecule in the small vibrations regime about its non-degenerate minimum, and shift the intramolecular coordinates to have $b=0$ at equilibrium, and then scale $b = \varepsilon_b \tilde b$ and $p_b = \varepsilon_b^{-1} \tilde p_b$.
As for the planar case, we assume that the potential scales to
\[ U(q;\varepsilon) = U_b (b) + \varepsilon_c^2 U_c^0 (x) + \varepsilon_c^4 U_c^2 (q; \varepsilon), \]
and then choose $\varepsilon_b$ such that
\[ U(q;\varepsilon) = \bar U_b^0 + \varepsilon_b^{-2} \sum_{i=1}^{3 n_b-6} \bar U_{b i}^2 b_i^2 + \varepsilon_c^2 U_c^0 (x) + \varepsilon_c^4 U_c^2 (q; \varepsilon) + \order{\varepsilon_b^5}. \]
That is, we are assuming that the molecule is strongly bonded, $ \del_{bb}^2 U_b (0) \sim \varepsilon_b^{-4} $.
Note that we have chosen normal mode intramolecular coordinates for which $\bar U_{\beta ij}^2 = \tilde U_{\beta i}^2 \delta_{ij}$.
The reduced kinetic and centrifugal energies contain both intermolecular and intramolecular terms and are scaled in Appendix \ref{spatialDetials}.
The Hamiltonian function becomes
\begin{align*}
H & \left( z; \lambda, \varepsilon \right) =
\frac{\varepsilon_c^{-2}}{2} \left( \varepsilon_b^{-2} \sum_{i=1}^{3n_b-6} \bar U_{b i}^2 ( p_{b i}^2 + b_i^2)
+  \sum_{i,j=1}^2 v_{\beta i} (z;\lambda,\varepsilon) J^{ij}_{\beta 0} (\beta) v_{\beta j} (z;\lambda,\varepsilon) 
+ I^{11}_0 (\beta) p_\lambda^2 \right) \\
&+ 
\frac{p_x^2}{2}   + \frac{1}{2} \sum_{i,j=1}^3 l_i (z_\lambda;\lambda) I^{ij}_2 (x,\beta)  l_j (z_\lambda;\lambda) + U_c^0 (x) - \frac{1}{2} \sum_{i,j=1}^2 v_{\beta  i} (z;\lambda,\varepsilon) J^{ij}_{\beta 2} (x,\beta) v_{\beta  j} (z;\lambda,\varepsilon) \\
&+ \frac{\varepsilon_c^2}{2}  \left( 
 \sum_{i,j=1}^2 v_{\beta  i} (z;\lambda,\varepsilon) J^{ij}_{\beta 4} (x,\beta) v_{\beta  j} (z;\lambda,\varepsilon) +  \sum_{i,j=1}^3 l_i (z_\lambda;\lambda) I^{ij}_4 (x,\beta)  l_j (z_\lambda;\lambda) + U_c^2 (q;0) \right)\\
 &+ \text{ h.o.t.},
\end{align*}
where $K^{-1}_\beta (q) = K_{\beta 0}^{-1} (\beta) - \varepsilon_c^2 J_2 (x,\beta) + \varepsilon_c^4 J_4 (x,\beta)$ and $I^{-1} (q) = I^{-1}_0 (\beta) + \varepsilon_c^2 I_2^{-1} (x,\beta) + \varepsilon_c^4 I_4^{-1} (x,\beta)$
for some $J_i$,
see Appendix \ref{spatialDetials}.
We are using angular momenta $l$ and the non-canonical momenta 
\[v_{\beta  i} (z;\lambda,\varepsilon) = p_{\beta i} - \sum_{j=1}^3 A_{\beta ij} (x,\beta; \varepsilon) l_j (z_\lambda;\lambda),\]
where $A_{\beta ij} (q) = (A_{\beta i1}(\beta), \varepsilon^2_c A_{\beta i2}(x,\beta),\varepsilon^2_c A_{\beta i3}(x,\beta))+ \cdots$, as place-holders.

\begin{figure}
\begin{center}
\begin{pspicture}(12,4.5)
		\rput(2.5,2.2){\includegraphics[width=0.30\textwidth]{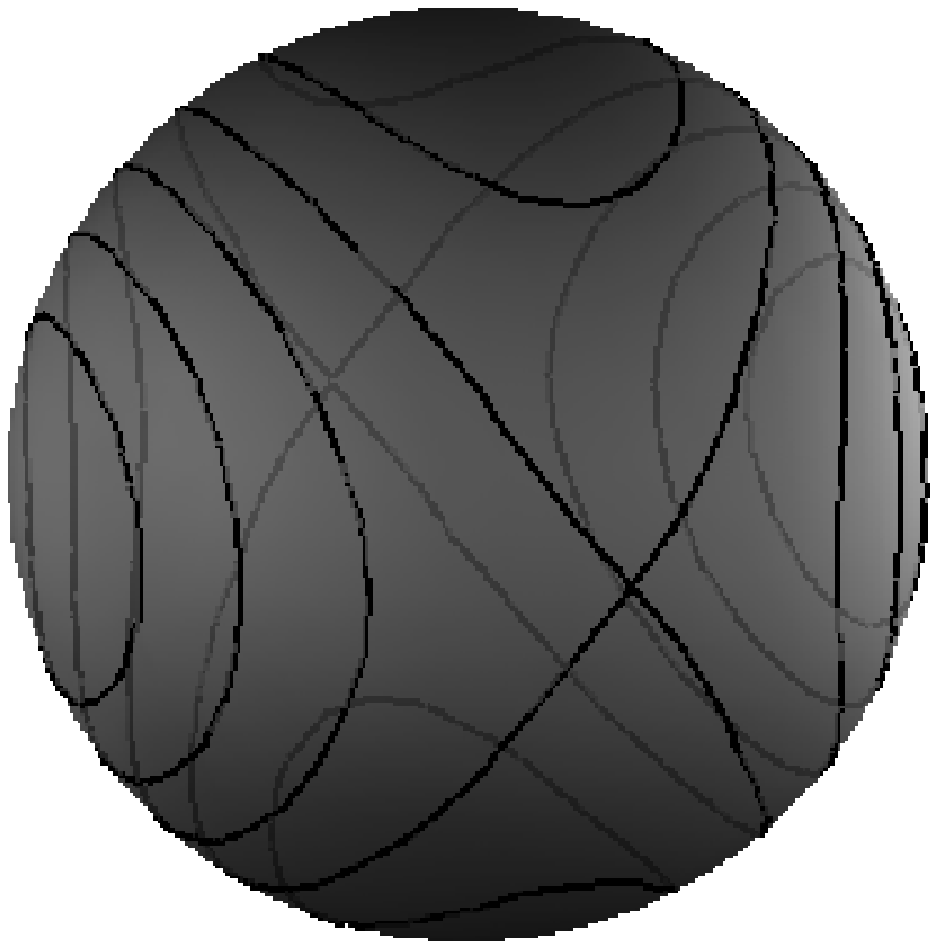}}
		\rput(9.5,2.2){\includegraphics[width=0.30\textwidth]{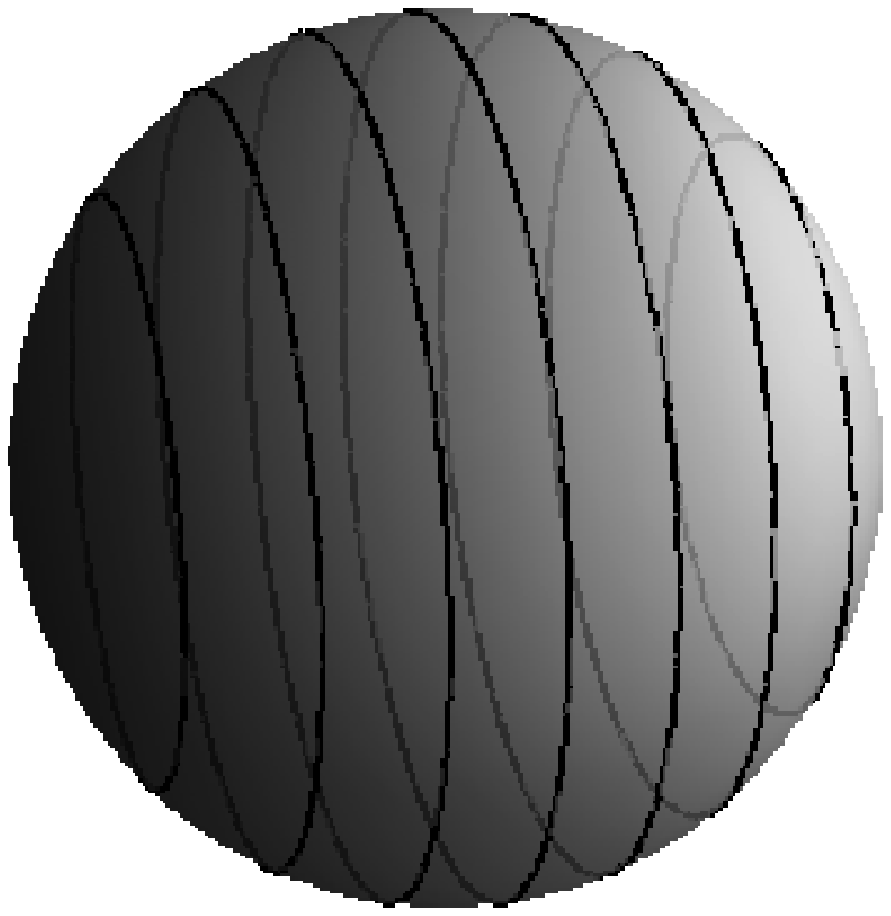}}
		\rput(3.3,1.12){{\white \bm $\bar z_\lambda^2$}}
		\rput(2.6,4.6){{\bm $\bar z_\lambda^3$}}
		\rput(0,2){{\bm $\bar z_\lambda^1$}}
		\rput(7.1,2){{\bm $\bar z_\lambda^1$}}
	\end{pspicture}
\end{center}
\caption{Angular momentum sphere with equipotential lines of the centrifugal energy, when the moment of inertia has three distinct principal moments (left), or two equal moments (right).}
\label{AMcentrif1}
\end{figure}

Next, we consider
the angular momentum \DoF $z_\lambda \in \Sbb^2_\lambda$. 
From the scaled Hamiltonian, we note that its dynamics are coupled to the internal one and that it appears in terms of orders both $\varepsilon_c^{-2}$ and $\varepsilon_c^0$.
In order to ensure that we have a normally hyperbolic capture \TM about $\bar x_c$, we need that the intermolecular distance degree of freedom $(x,p_x)$ is more hyperbolic than the angular momentum degree of freedom. We will therefore find the necessary scaling for this to be the case and apply it.

Considering the atom and the molecule as constituting a single ``body'', 
if the molecule is in equilibrium with itself and with respect to the atom, we obtain a rigid body and the Hamiltonian reduces to the centrifugal terms. 
Rigid bodies follow closed curves on the angular momentum sphere with equilibrium points when $l$ is parallel to the eigenvectors of $I^{-1}$, called \emph{principal axes} \cite{Deprit1967b}.
Typical rigid body dynamics for the case of distinct eigenvalues, or \emph{principal moments}, is depicted in Figure \ref{AMcentrif1}.
In general, the angular momentum \DoF doesn't follow closed curves on $\Sbb^2_\lambda$, since it is coupled with the internal ``deformation'' dynamics. 
The equilibria of the system occur when $q$ is a critical point of the effective potential $V$, $v = 0$, and again $l$ is parallel to the principal axes.
We therefore consider the centrifugal energy
\[ E_\lambda = \frac{1}{2} l^T (z_\lambda; \lambda) I^{-1} (q; \varepsilon) l (z_\lambda; \lambda), \]
which is not conserved, in more detail.

First, we consider the moment of inertia tensor and its principal moments and axes.
The eigenvectors $\eta_i$ of $I^{-1}$ are the same as those of $I$, whereas the eigenvalues $\mu_i$ are the reciprocals.
$I$ has real eigenvalues, since it is a real positive-definite matrix, and if these are distinct then the eigenvectors are orthogonal.
To order $\varepsilon_b^0$, the eigenvalues and eigenvectors of $I^{-1}$ are
\begin{align*}
\mu_1 (q) & = I^{11}_\beta (\beta) + \cdots, & \eta_1 (q) &= x_1 + \order{\varepsilon^2_c}, \\
 \mu_i (q) & = \varepsilon^2_c x^{-2} + \varepsilon^4_c x^{-4} \mu_{i4} (\beta) + \cdots, & \eta_i (q) &\in \lbrace x_2, x_3 \rbrace + \order{\varepsilon^2_c},
\end{align*}
where
\[
\mu_{i4} (\beta) = \frac{1}{2} (I_{22}^\beta (\beta) + I_{33}^\beta (\beta) ) \pm \sqrt{\frac{1}{4} (I_{22}^\beta (\beta) - I_{33}^\beta (\beta) )^2 + I_{23}^\beta (\beta)^2 },
\]
and $ i = 2,3$.
Thus in order to have distinct principal moments and axes, we require either $I_{22}^\beta (\beta) \neq I_{33}^\beta (\beta)$, or $I_{23}^\beta (\beta) \neq 0$.
For our 3D molecule $B$ with three distinct moments,
as it rotates relative to the distant atom, we expect to find three pairs of points (on the attitude sphere $\Sbb^2_B$) at which two of the moments of $I(q)$, and so $I^{-1}(q)$, are non-distinct.

Then, we consider the actual centrifugal energy and its critical points. Given a fixed configuration $q$ with distinct $\mu_1 (q) > \mu_2 (q) > \mu_3 (q)$, 
we write
\[ l (q, Z_\lambda; \lambda) = P_\lambda \eta_1 (q) + \sqrt{\lambda^2 - P_\lambda^2} \sin Q_\lambda \eta_2 (q) + \sqrt{\lambda^2 - P_\lambda^2} \cos Q_\lambda \eta_3 (q), \]
\ie~consider Serret-Andoyer coordinates $Z_\lambda$ obtained by projecting onto the principal axes, cf.~Appendix \ref{biGauge}.
Then
\[ E_\lambda = \frac{1}{2} \left( \mu_1 (q) P_\lambda^2 + \mu_2 (q) (\lambda^2 - P_\lambda^2) \sin^2 Q_\lambda  + \mu_3 (q) (\lambda^2 - P_\lambda^2) \cos^2 Q_\lambda \right), \]
so the critical points are $\bar Z_\lambda^2 = (0,0)$, $(0,\pi)$, $\bar Z_\lambda^3 = (\frac{\pi}{2},0)$, $ (\frac{3 \pi}{2},0)$, and $\bar Z_\lambda^1 = (Q_\lambda,\pm \lambda)$.
The superscript denotes the principal axis to which $l$ is parallel. 
For non-distinct eigenvalues $\mu_2 (q ) = \mu_3 (q)$, choosing generalised eigenvectors for
$\eta_2, \eta_3$, gives
\[ E_\lambda = \frac{1}{2} ( ( \mu_1 (q) - 2 \mu_2 (q) ) P_\lambda^2 + 2 \mu_2 (q) \lambda^2, \]
so the critical points are the degenerate $(Q_\lambda, 0)$, and $(Q_\lambda, \pm \lambda)$. This is depicted in Figure \ref{AMcentrif1}.

For arbitrary $\lambda \sim 1$, in order to ensure that the centrifugal terms are higher order than the intermolecular $(x,p_x)$ ones, we restrict our attention to energies just above the critical
\[\bar E_\lambda^2 = \frac{\mu_2}{2} \lambda^2 = \frac{\lambda^2}{2} (\varepsilon^2_c x^{-2} + \varepsilon^4_c x^{-4} \mu_{24} (\beta)) + \cdots \sim \varepsilon_c^2 \lambda^2 \]
such that the angular momentum \DoF is confined to a small annulus $\Abb^2_{\lambda}$ that doesn't contain $\bar Z^1_\lambda$. 
If all the energy of the system is in $E_\lambda$, then for the non-distinct case we have $P_\lambda = 0$ at $E_\lambda = \bar E_\lambda^2$, whereas for the distinct case
\[P_\lambda^2 (0,\bar E_\lambda^2 ) = \frac{\varepsilon_c^4 \lambda^2 (\mu_{24} - \mu_{34})}{x^4 \mu_1 } + \cdots \sim \varepsilon_c^4 \lambda^2.\]
Furthermore, $p_\lambda = P_\lambda + \order{\varepsilon_c^2}$, so bounding
$E < \bar E_\lambda^2 + \Delta$, with $\Delta$ small, gives $p_\lambda = 0 + \order{\varepsilon_c^2} + \order{\varepsilon^4_c}$.
Thus, we scale $p_\lambda = \varepsilon_c^2 \tilde p_\lambda$, giving
\[ l = \lambda (0, \sin q_\lambda, \cos q_\lambda) + \varepsilon_c^2 p_\lambda (1,0,0) + \order{\varepsilon_c^4}.\]

Next, we consider the rotational degree of freedom, with non-canonical rotational momenta
\[v_{\beta  i} (z ;\lambda,\varepsilon) = p_{\beta i} - \varepsilon_c^2 \left( A_{\beta i1} (\beta) p_\lambda + A_{\beta i2} (x,\beta) \lambda \sin q_\lambda + A_{\beta i3} (x,\beta) \lambda \cos q_\lambda \right) + \cdots \]
and ensure that the $(x,p_x)$ dynamics dominates this one also.
Even though we have removed the coupling of vibrations and rotations up to order $\varepsilon^0$, $p_{\beta i}$ are not conserved since the reduced metric $K$ is a function of $\beta$. However, the rate of change of $p_\beta$ is a function of $p_\beta^2$ up to order $\varepsilon_c^2$, so we consider a molecule that is initially rotating slowly, which will take a long time to increases its rotational velocity.
Thus, we scale $p_{\beta i} = \varepsilon_c^2 \tilde p_{\beta i}$, so $v_{\beta i} \sim \varepsilon_c^2$. 

Finally 
\begin{align*}
H \left( z; \lambda, \varepsilon \right)  &=
\varepsilon_c^{-2} \varepsilon_b^{-2} \sum_{i=1}^{3n_b-6} \frac{\bar U_{b i}^2}{2} ( p_{b i}^2 + b_i^2) + 
\frac{1}{2} p_x^2  + \frac{\lambda^2}{2 x^2} + U_c^0 (x) \\
& + \varepsilon_c^2  \left( 
\frac{1}{2} \sum_{i,j=1}^2 v_{\beta i} (z ;\lambda,0) J^{ij}_{\beta 0} (\beta) v_{\beta j} (z ;\lambda,0) + \frac{1}{2} I^{11}_0 (\beta) p_\lambda^2 \right. \\
& + \left.   \sum_{j=2}^3 p_\lambda I^{1j}_2 (x,\beta) l_j^0 (q_\lambda;\lambda) + \frac{1}{2} \sum_{i,j=2}^3 l_i^0 (q_\lambda;\lambda) I^{ij}_4 (x,\beta)  l_j^0 (q_\lambda;\lambda) + U_c^2 (q;0) \right) + \order{\varepsilon_c^4,\varepsilon_b^1},
\end{align*}
and
\[ \omega = \sum_{i=1}^{3n_b-6} \rmd {b_i} \wedge \rmd p_{b i} + \varepsilon_c^2 \sum_{i=1}^{2} \rmd \beta_i \wedge \rmd p_{\beta i} 
+ \varepsilon_c^2 \rmd q_\lambda \wedge \rmd p_\lambda + \rmd x \wedge \rmd p_x,\]
giving the equations of motion, up to order $\varepsilon^0$, as
\begin{align*}
\dot b_i &= \varepsilon^{-2} \varepsilon_b^{-2} \bar U_{bi}^2 p_{b i}, &
\dot \beta_i &= \del_{p_{\beta i}} H_2 (z;\lambda) , & \dot q_\lambda &= \del_{p_\lambda} H_2 (z;\lambda), & \dot x &= p_x, \\
\dot p_b &= - \varepsilon^{-2} \varepsilon_b^{-2} \bar U_{bi}^2 b_i , &
\dot p_{\beta i} &= - \del_{{\beta_i}} H_2 (z;\lambda) , & \dot p_\lambda &= - \del_{q_\lambda} H_2 (z;\lambda), & \dot p_x &= - \del_x V_c^0 (x; \lambda), 
\end{align*}
where $ H_2 (z;\lambda) \sim \varepsilon_c^2$. Thus, about the maximum $\bar x_c$, 
the $(x,p_x)$ \DoF is more hyperbolic than both the attitude $(\beta, p_\beta)$ and the angular momentum $z_\lambda$ ones. The submanifold
\[N_0 = \lbrace z \in M_\lambda \vert x = \bar x_c (\lambda), \; p_x = 0 \rbrace \]
is therefore invariant to order $\varepsilon^0$, and normally hyperbolic.
Taking $N_0$ as an approximation to the true normally hyperbolic submanifold $N$ nearby, and 
\[S_0 = \lbrace z \in M_\lambda \vert x = \bar x_c ( \lambda) \rbrace,\]
as an approximate dividing manifold, we can find the restricted Hamiltonian functions
\begin{align*}
H_N \left( z; \lambda, \varepsilon \right) &=
\varepsilon_c^{-2} \varepsilon_b^{-2} \sum_{i=1}^{3n_b-6} \frac{\bar U_{b i}^2}{2} ( p_{b i}^2 + b_i^2 ) + \varepsilon_c^2 \bigg( 
\frac{1}{2} \sum_{i,j=1}^3 v_{\beta i} (z ;\lambda,0) G^{ij}_{\beta 0} (\beta) v_{\beta j} (z ;\lambda,0) \\
& + \frac{1}{2} I^{11}_0 (\beta) p_\lambda^2  + \sum_{j=2}^3 p_\lambda I^{1j}_2 (\bar x_c,\beta) l_j^0 (q_\lambda;\lambda) + \frac{1}{2} \sum_{i,j=2}^3 l_i^0 (q_\lambda;\lambda) I^{ij}_4 (\bar x_c,\beta)  l_j^0 (q_\lambda;\lambda) \\
&+  U_c^2 (q;0) \bigg) + \order{\varepsilon_c^4,\varepsilon_b^1},
\end{align*}
modulo constant terms, and $H_S$ to leading orders. 
These give the \TSs $N_E = H_N^{-1}(E)$ and dividing surfaces $S_E = H_S^{-1}(E)$, respectively.

As for the planar examples, it is simpler to study the Morse bifurcations if we minimise the reduced Hamiltonians over the positive-definite coordinates, namely $b, p_b$, $ v_\beta$
and $p_\lambda$. 
This can be simplified by using canonical angular momentum coordinates $Z_\lambda$ aligned with the principal axes, and setting $b = p_b = v_\beta = P_\lambda = 0$ in $H_N$, to obtain
\begin{align*}
V_N^c  \left( \beta, Q_\lambda; \lambda, \varepsilon \right) 
&: = H_N \left( \bar x_c (\lambda), 0 , \beta, Q_\lambda, 0, 0, 0, 0 ; \lambda, \varepsilon \right)\\
&= \varepsilon_c^2 \left( \frac{\lambda^2}{2}  \left( \mu_{24} (\beta ) \sin^2 Q_\lambda + \mu_{34} (\beta ) \cos^2 Q_\lambda \right) + \bar U_c^2 (\beta ;0) \right) + \order{\varepsilon_c^4,\varepsilon_b^1}.
\end{align*}
We are then interested in the level-sets of $V_N^c$ and their bifurcations. 
The domain of $V_N^c$ is a subset of $N$, which is codimension-2 in the reduced state space $\tilde M_\lambda$. The latter is some $\Sbb^2_\lambda$ fibre bundle over $T^* ( Q_{I_d}/SO(3) )$, so also $N$ and $S$ could in general be non-trivial bundles.
We shall restrict our attention to subsets for which the bundle is trivial.
Furthermore, we are considering energies below that at which the molecule dissociates, and up to just above the centrifugal energy for the angular momentum aligned with the $\eta_2 (\beta) $ principal axis. 

The critical points $(\bar \beta, \bar Q_\lambda)$ of the frozen, restricted effective potential $V_N^c$ are given by
\begin{align*}
(\mu_{24} (\bar \beta) - \mu_{34} (\bar \beta) ) \sin \bar Q_\lambda \cos\bar Q_\lambda &= 0, \\
\frac{\lambda^2}{2}  \left( \del_\beta \mu_{24} (\beta ) \sin^2 Q_\lambda + \del_\beta \mu_{34} (\beta ) \cos^2 Q_\lambda \right) + \del_\beta \bar U_c^2 (\beta ;0) &= 0.
\end{align*}
The first equation is satisfied trivially for $\hat \beta$ at which the two principal moments are equal $\mu_{24} (\hat \beta) = \mu_{34} (\hat \beta)$.
We shall consider examples of $V_N^c$ that are Morse functions, \ie~have non-degenerate critical points $(\bar \beta, \bar Q_\lambda)$ with $\bar \beta \neq \hat \beta$, satisfying either  
\[
\bar Q_\lambda^3 =k \pi \quad \text{  and  } \quad  \del_\beta \left(\frac{\lambda^2}{2}  \mu_{34} + U_c^2 \right) (\bar \beta) = 0,
\]
or
\[
\bar Q_\lambda^2 = k \pi + \frac{\pi}{2} \quad \text{  and  } \quad \del_\beta \left(\frac{\lambda^2}{2} \mu_{24} + U_c^2 \right) (\bar \beta) = 0,
\]
for $k \in \Zbb$, cf.~\cite[Section IV.E]{Littlejohn1997}.
Furthermore, the Morse function $V_N^c$ has at least two non-degenerate minima at $(\bar \beta^0, \bar Q_\lambda^3)$ due to the symmetry of the centrifugal terms.

The sequence of Morse bifurcations of the level sets of $V_N^c$, and therefore of $N_E$ and $S_E$, depends on the relative size of the centrifugal and the reduced potential $U_c^2$ energies.
This will determine the relation of the different critical energies. Critical points with the same attitude $\beta$ but the angular momentum aligned with different principal axes have energies that differ by $\lambda^2$, whereas the difference in energy for different attitudes depends on the atom-molecule pair.

The simplest case is when the first Morse bifurcation encountered as the energy is increased from the minima involves the angular momentum angle, and the system goes from rotating about the $\eta_3 (\bar q)$ axis to rotating more freely about $\eta_2 (\bar q)$ as well.
This bifurcation occurs at the critical energy for the $(\bar \beta^0, \bar Q_\lambda^2)$ critical points.
In this case both the domain of $V_N^c$ and the subsets of $N$ and $S$ of interest are bundles over a contractible base space and so trivial \cite[Theorem 11.6]{Steenrod1951}.
The frozen energy levels $\tilde N_{\leq E}$ bifurcate from $\Sbb^0 \times \Bmbb^3$ to $ \Sbb^1 \times \Bmbb^2$, so the \TSs $N_E$ go from $\Sbb^0 \times \Sbb^{6 n_b - 7} $ to $\Sbb^1 \times \Sbb^{6 n_b - 8}$, and similarly the dividing surfaces.

\begin{figure}
\begin{center}
\begin{pspicture}(12,11.5)
		\rput(2.5,8.5){\includegraphics[width=0.33\textwidth]{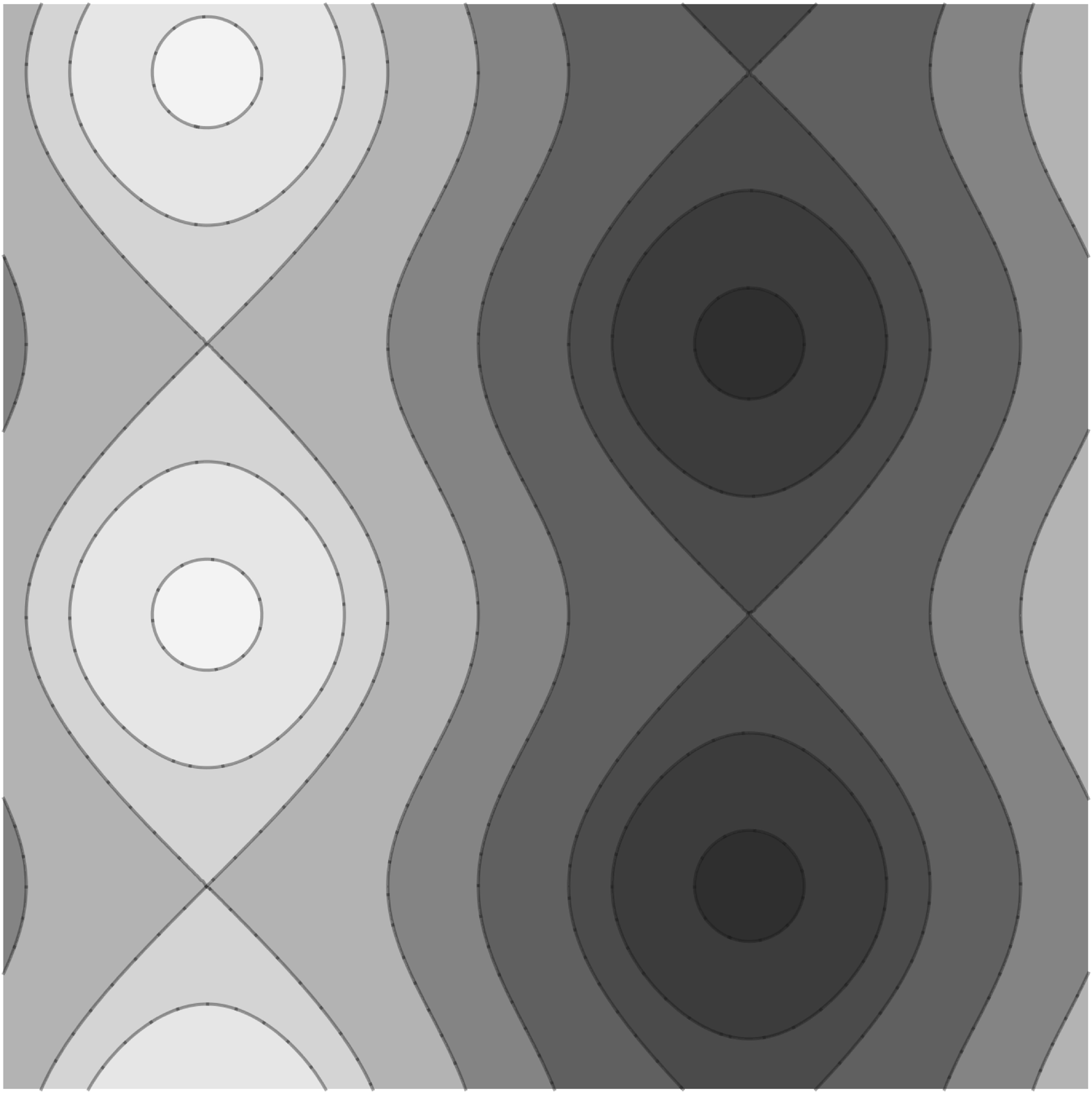}}
		\rput(9.5,8.5){\includegraphics[width=0.33\textwidth]{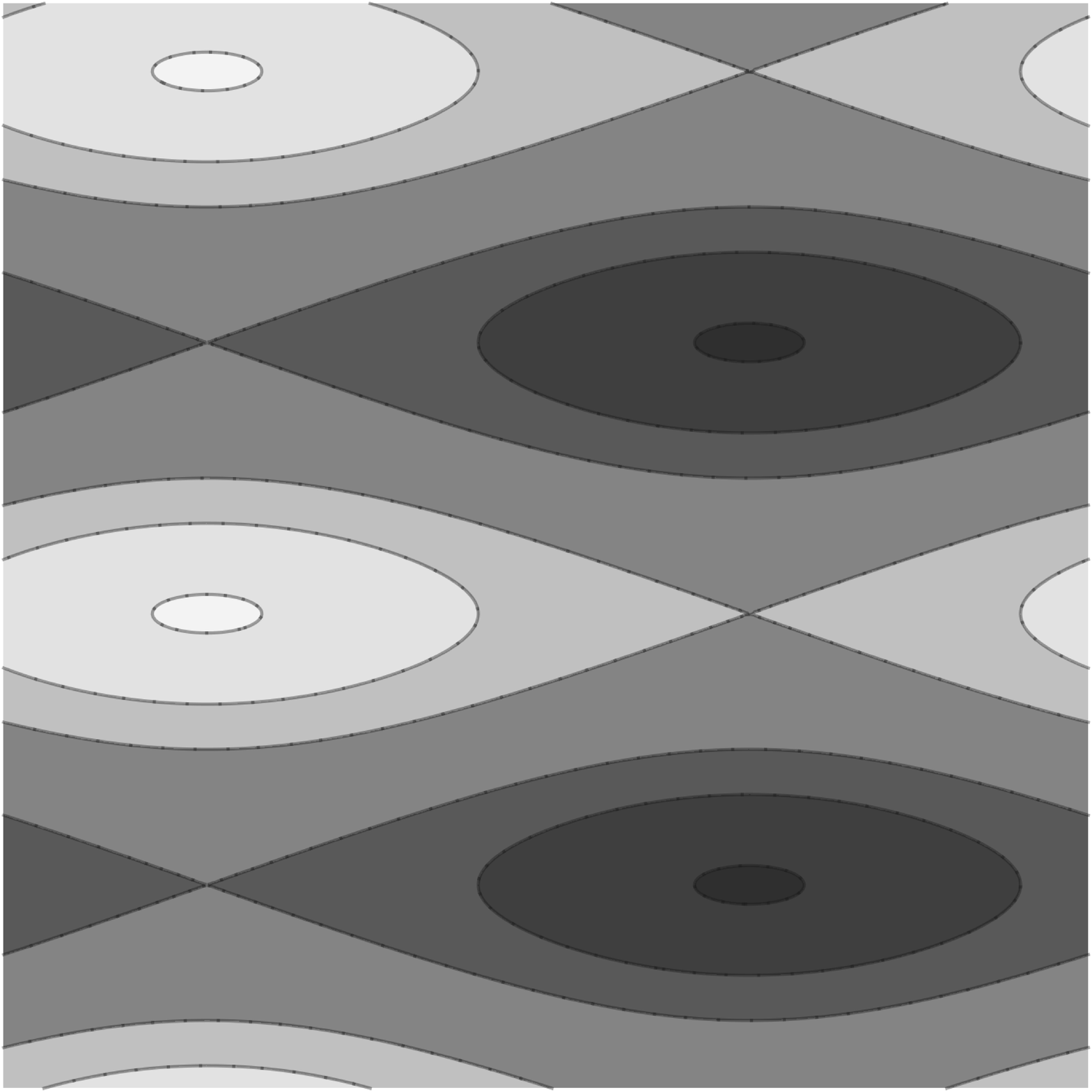}}
		\rput(12.2,5.65){$\beta_1$}
		\rput(6.6,11.2){$Q_\lambda$}
		\rput(-0.4,11.2){$Q_\lambda$}
		\rput(-0.4,6.9){$\bar Q_\lambda^3$}
		\rput(-0.4,8.2){$\bar Q_\lambda^2$}
		\rput(-0.7,9.5){$\bar Q_\lambda^3 + \pi$}
		\rput(-0.7,10.7){$\bar Q_\lambda^2 + \pi$}
		\rput(5.2,5.65){$\beta_1$}
		\rput(3.4,5.65){$\bar \beta_1^0$}
		\rput(0.9,5.65){$\bar \beta_1^1$}
		\rput(2.5,2.5){\includegraphics[width=0.33\textwidth]{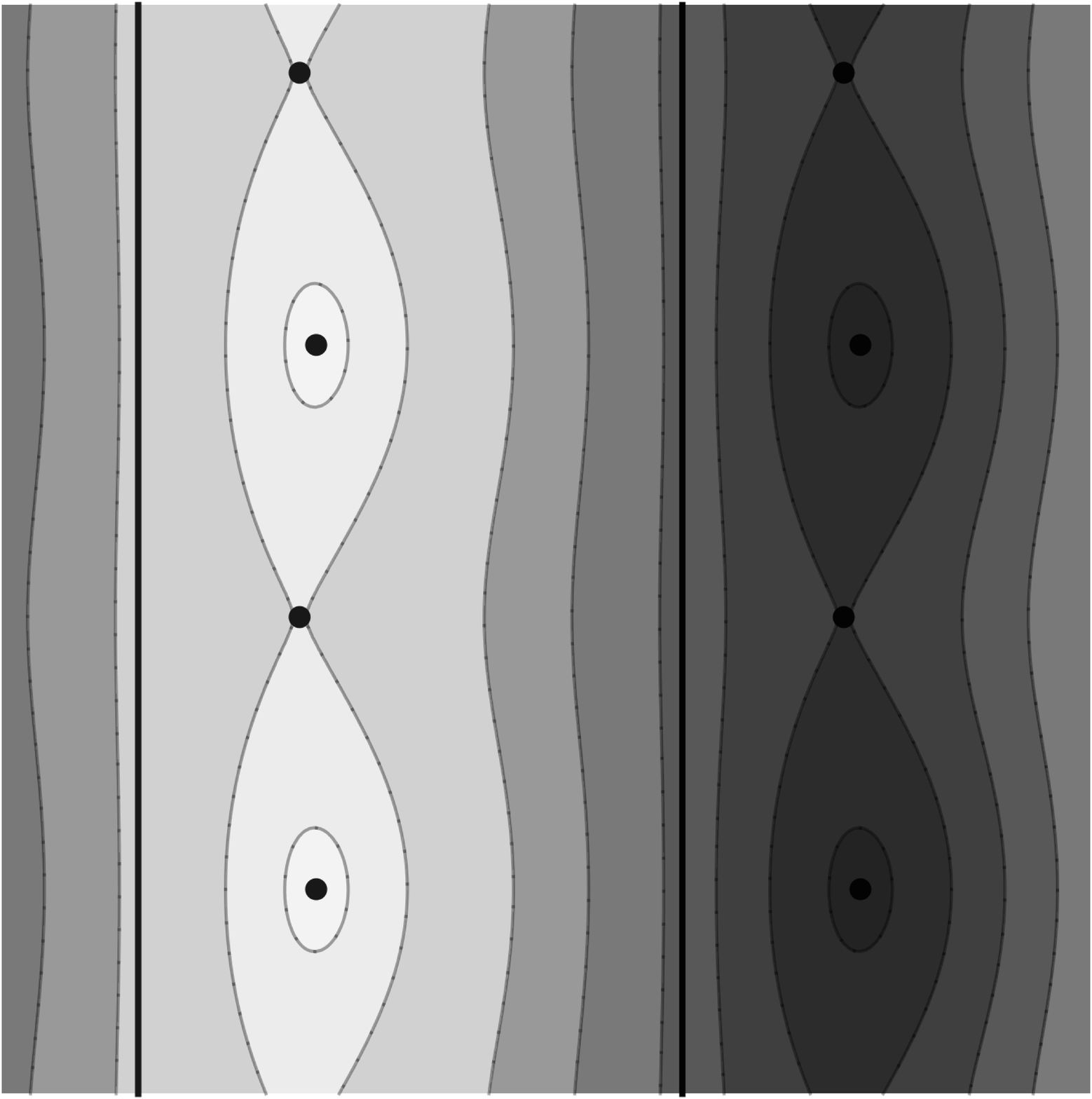}}
		\rput(9.5,2.5){\includegraphics[width=0.33\textwidth]{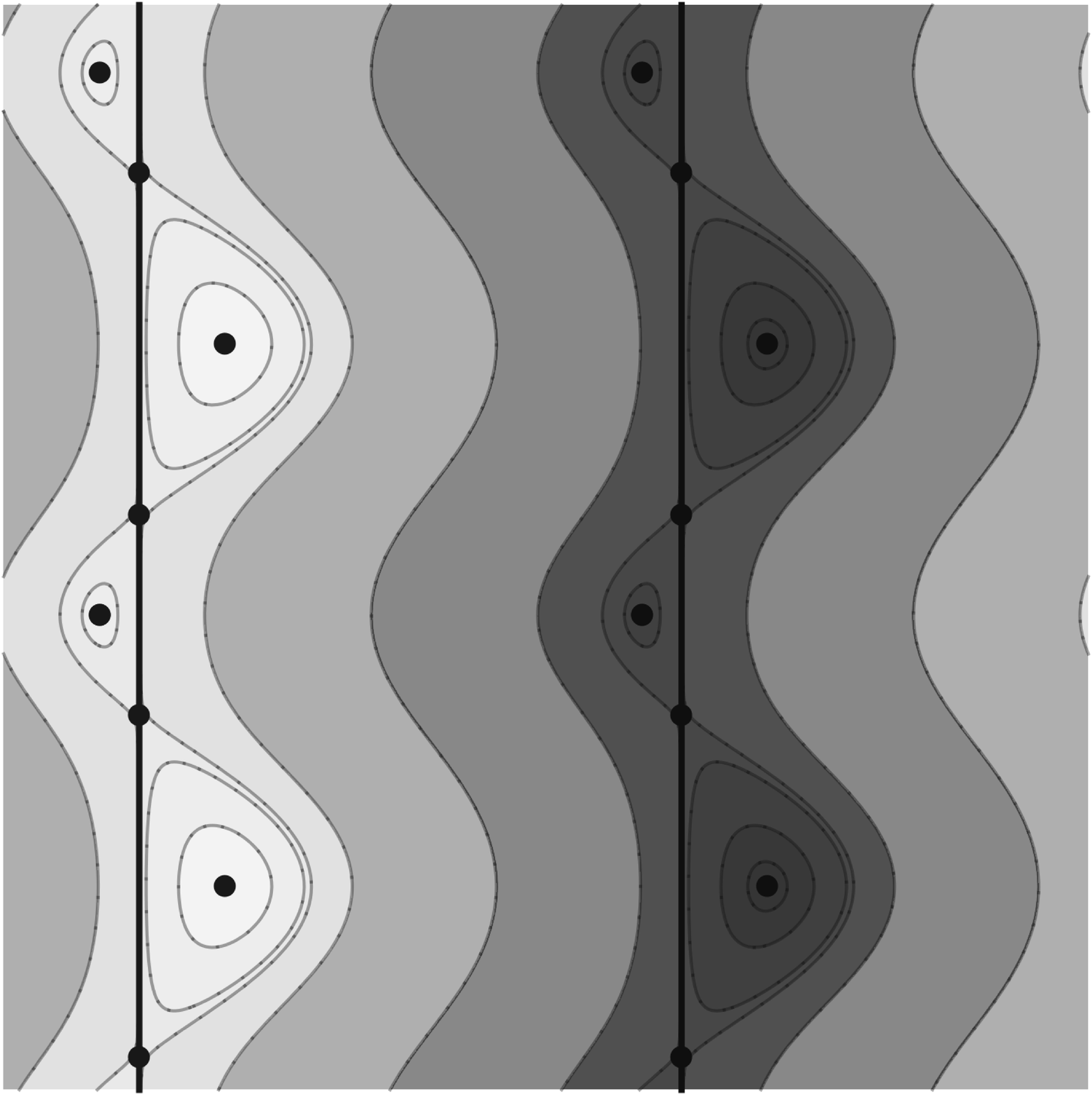}}
		\rput(12.2,-0.3){$\beta_1$}
		\rput(6.6,5.2){$Q_\lambda$}
		\rput(-0.4,5.2){$Q_\lambda$}
		\rput(5.2,-0.3){$\beta_1$}
		\rput(0.6,-0.3){$\hat \beta_1$}
		\rput(3.5,-0.3){$\hat \beta_1 + \pi$}
	\end{pspicture}
\end{center} 
\caption{Contour plots of example functions on the $(\beta_1, Q_\lambda)$ torus, where darker regions represent lower energies. Top row: for no values $\hat \beta_1$ with non-distinct principal moments. Bottom row: for non-distinct principal moments $\mu_2$ and $\mu_3$ at $\hat \beta_1$ depicted by vertical back lines. Case in which the value of the function at $(\bar \beta^0_1, \bar Q_\lambda^2)$ is smaller than that at $(\bar \beta^1_1, \bar Q_\lambda^3)$ on the left, and vice-versa on the right.}
\label{spatialCase1}
\end{figure}

As the energy is increased further, 
we will reach critical values at which also the attitude coordinates are involved in Morse bifurcations.
We will consider the case in which the energy does not change significantly as the molecule rotates in one direction, with respect to the atom, but does when it tries to rotate in the other direction
and a potential $U_c^2$ on $\Sbb^2$ that has a minimum $\bar \beta^0$, a saddle $\bar \beta^1$ and two maxima $\bar \beta^2$, restricting our attention to the annulus $\Abb^2 \subset \Sbb^2$ containing $\bar \beta^0$ and $\bar \beta^1$. Therefore, choosing the angles appropriately, only one is involved in Morse bifurcations, whereas the other contributes positive definite terms. 
The subset of the \TM $N$ of interest is a bundle over $\Sbb^1 \times \Bmbb^{6 n_b -9}$ which we claim is trivial. Firstly, we note that it is equivalent to the product of a bundle over $\Sbb^1$ with $\Bmbb^{6 n_b -9}$ via homotopy-type arguments \cite[Theorem 11.4]{Steenrod1951}, cf.~bundles over contractible spaces being trivial.
The characterisation of bundles over spheres with structure group $G$ depends on certain homotopy groups of $G$ \cite[Theorem 18.5]{Steenrod1951}. Our fibres are diffeomorphic to $\Sbb^2$, or subsets of it, and the diffeomorphism group of $\Sbb^2$ is the orthogonal group $O(3)$.
However, $N$ is orientable so both elements of the product must be orientable. Thus, given that the bundle over the circle is orientable, we restrict our attention to the orientation preserving diffeomorphisms $SO(3)$ and find that the bundle over the circle is a product, and therefore our original bundle is also trivial \cite[Section 26]{Steenrod1951}.
Note however that not all orientable surface bundles over the circle are product bundles, as we can construct non-trivial bundles with fibres diffeomorphic to $\Tbb^2$, for example.

$V_N^c$ can be minimised over the irrelevant attitude to obtain a function on the torus $\Tbb^2$ for $(\beta_1, Q_\lambda)$, say. 
There are two possible scenarios for this case, the first is that $\Tbb^2$ does not contain points $\hat \beta$ at which the $\mu_2$, $\mu_3$ principal moments become equal.
The order of the bifurcations then depends on the relative heights of the critical energies, and both cases are straightforward, see Figure \ref{spatialCase1}.
The other scenario is when $\Tbb^2$ does contain $\hat \beta$. We shall consider the case in which it contains only one pair of such points. 
Contour plots for the restricted function on $\Tbb^2$ are given in Figure \ref{spatialCase1}.
If the centrifugal energy is smaller than the attitude potential, then the points $\hat \beta_1$ at which the moments $\mu_2$, $\mu_3$ are not distinct do not play a role in the Morse bifurcations, which are the same as those for the case when $\Tbb^2$ does not contain $\hat \beta$, as we can see by comparing the left hand side of Figures \ref{spatialCase1}.
Instead, when the molecular potential is smaller than the centrifugal one, depicted on the right in Figure \ref{spatialCase1}, we see that the points $\hat \beta_1$ do play a significant role in the bifurcations and the sub-level sets of the torus bifurcate as follows
\[ \Sbb^0 \times \Bmbb^2 \text{ to } \Sbb^0 \times \Sbb^0 \times \Bmbb^2 \text{ to } \Sbb^1 \times \Bmbb^1 \text{ to } X^c \text{ to } Y^c \text{ to } \Tbb^2,  \]
where $X^c$ and $Y^c$ can be written as handlebodies using Morse Theorem B \cite{MacKay2014}.
Therefore, the sub-level sets of the capture \TM $N_{\leq E}$ have the following sequence of bifurcations
\[ \Sbb^0 \times \Bmbb^{6 n_b - 6} \text{ to } \Sbb^0 \times \Sbb^0 \times \Bmbb^{6 n_b - 6} \text{ to } \Sbb^1 \times \Bmbb^{6 n_b - 7} \text{ to } X \text{ to } Y \text{ to } \Tbb^2 \times \Bmbb^{6 n_b - 8},  \]
and the \TSs
\[ \Sbb^0 \times \Sbb^{6 n_b - 7} \text{ to } \Sbb^0 \times \Sbb^0 \times \Sbb^{6 n_b - 7} \text{ to } \Sbb^1 \times \Sbb^{6 n_b - 8} \text{ to } \del X \text{ to } \del Y \text{ to } \Tbb^2 \times \Sbb^{6 n_b - 9}. \]
Similarly for the dividing surfaces.

Finally, if we were to consider higher energies for this choice of $U_c^2$, the other attitude would also become involved in Morse bifurcations. Here again the $\hat \beta$ points would most likely lead to interesting sequences of Morse bifurcations, however we would also have to deal with the non-trivial nature of the fibre bundle. After the Morse bifurcations at the index-2 critical points $\bar \beta^2$, the base space would contain a 2-sphere, and many examples of non-trivial orientable bundles over these can be found. 
Thus before we can consider the full sequence of Morse bifurcations of the \DSs and \TSs and the transport for a larger range of energies, the bundle class of the reduced state space needs to be understood.


\section{Conclusions and discussion}
\label{conclude}

The purpose of this article was to show that
Morse bifurcations must be considered when studying transport problems for larger ranges of energies,
and more specifically to show the existence of Morse bifurcations of capture transition states and \DSs for bimolecular reactions.
By considering the different sequences of Morse bifurcations we were able to find interesting new transition states and dividing surfaces for general reactions with non-zero angular momentum, thus extending the dividing surface approach beyond the well known basic transport scenario. Other choices of molecules than those considered here will have similar capture transport problems and therefore similar transition states and bifurcations.

The flux of ergode through a dividing surface as a function of energy and the effect of the Morse bifurcations was considered in \cite{MacKay2014} and commented upon in Section \ref{atomDiatom} for planar atom-diatom reactions.
Seeing as the systems representing these examples have more than 2 degrees of freedom, apart from the unrealistic planar atom-frozen diatom case, the Morse bifurcations do not have a significant effect on the flux, which varies $C^{m-2}$ smoothly through these.

The actual use of capture rates as bounds on reaction rates is debatable, but largely depends on the reaction being considered.
However, these were only chosen to provide relatively simple examples of Morse bifurcations, and to show the importance of the attitude and angular momentum \DoFs in bimolecular reactions.

We considered bimolecular reactions with Euclidean symmetry and reduced them accordingly. 
Even though (symplectic) reduction theory is a highly developed subject, we faced a number of difficulties when considering these examples. 
Setting aside the fact that singular reduction was required, due to the nature of the rotational symmetry, and that singular cotangent bundle reduction is still not a complete theory, there is a large gap between the reduction theory literature and applications.
Even if one restricts one's attention to the principal non-singular stratum, it is not an easy task to find suitable charts.
Some of the literature avoids charts altogether, focusing instead on the global geometric properties of the reduced spaces, whereas the celestial mechanics literature considers charts for different regions of the reduced space.
The most common approach in the molecular literature is to restrict one's attention to non-collinear configurations such that the gauge theoretic approach to cotangent bundle reduction provides a set of charts, as reviewed in Appendix \ref{biGauge}.
However, here we face the opposite issue, namely the reduced space is an $\Sbb^2$ fibre bundle, due to the angular momentum degree of freedom, but the global nature of this bundle is generally not discussed in the literature. 
We feel that more work is needed, both on charts for the reduced spaces and on their global nature, and that this would improve our understanding of molecular reactions, and also other $n$-body systems.

By considering normal molecules, with a fixed equilibrium configuration and energies below that at which either of the two molecules dissociates, collinearity becomes a decreasing concern with increasing size of the molecules, namely codimension-$(2n_i-5)$ where $n_i \geq 3$ is the number of atoms in the $i$th molecule ($i = 1,2$), and the chemistry of the molecule is not taken into account. 
However, for smaller molecules, higher energies, or other transport problems we may need to consider collinear configurations.
For non-zero angular momentum, collinear configurations are a subset of the principal reduced stratum.
However due to collinear configurations having non-trivial configuration space isotropy, we cannot find charts via the gauge theoretic approach to cotangent bundle reduction.
The issue is therefore not one of reduction per se, but only of finding suitable coordinates.
The transport problem and bifurcations of transition states will be the same as those considered in Section \ref{spatialBimol}.
For more than seventy years, chemists have been using charts obtained by modifying gauge theoretic cotangent bundle reduction \cite{Sayvetz1939}.
The idea is to pass to a rotating frame in which the collinear (equilibrium) configuration is along a chosen axis, say the $x_1$-axis, but retain the remaining rotational symmetry (about $x_1$) as an internal coordinate. Then by choosing the Eckart convention and the non-gauge invariant form of the kinetic energy, we find that the Lagrangian is not a function of the angular velocity about the collinear axis $\omega_1$, so we can obtain a Hamiltonian that is not a function of the first angular momentum component $l_1$. That is, $l_1$ is replaced by the canonical momentum conjugate to the ``internal'' rotation about the $x_1$-axis. These charts were first considered by Sayvetz \cite{Sayvetz1939}, though nowadays they are often attributed to Watson \cite{Watson1970}.
This procedure can be justified geometrically by applying the slice theorem (see e.g.~\cite[Section 2.3.14]{Ortega2004}) to configuration space in a neighbourhood of the collinear configurations, and then lifting the charts obtained to the cotangent bundle \cite{Roberts2006}.
Actually, with this understanding, charts can be obtained that are not those of the Eckart convention, i.e.~other gauges and internal coordinates.
This was used in examples by Kozin et al.~\cite{Kozin2000}.
Note that this is just the splitting of coordinates into internal coordinates and rotations and not an actual reduction, cf.~Appendix \ref{biGauge}. With these charts, we cannot simply pass to Serret-Andoyer coordinates to reduce the symmetry, seeing as the Hamiltonian is not a function of $l_1$. This is generally not addressed in literature.

The transport problems associated with reaction are usually a lot more complicated than the capture ones considered here, so their transition states may undergo a number of different bifurcations.
One simple example which should display much the same bifurcations as those seen here is isomerization reactions involving only one molecule. These will be the topic of a future publication.

We have concentrated here on how the transition state and dividing surface vary with energy, but 
the exact dependence of the capture transition states on the angular momentum and the possible loss of normal hyperbolicity for large values should be considered in detail.
Due to the high \DoFs of these systems, this is not a straightforward task.

It is also interesting to ask whether the dividing surface method can be extended to consider 
reactions with a varying external field or laser pulse, which need to be modelled as a non-autonomous Hamiltonian system;
or reactions that are not in the (dilute) gas phase, for which the product kinetic approximation leading to a low dimensional Hamiltonian system is not valid;
or even reactions out of equilibrium. 
Reviews of the dividing surface approach applied to the basic transport scenario of flux over a saddle for non-autonomous systems and Langevin systems can be found in \cite{Bartsch2008} and \cite{Kawai2011}, respectively. However, more work is required to consider general transport scenarios and fully understand transport in these systems.


\section*{Acknowledgements}
\addcontentsline{toc}{section}{Acknowledgements}

We thank Mark Roberts, Robert Littlejohn and Kevin Mitchell for pointing us to \cite{Roberts2006} and \cite{Littlejohn2002}, respectively, and two anonymous referees for comments and suggestions leading to improvements to the article. Strub was supported by an EPSRC studentship.


\appendix


\section{Charts for reduced $n$-body systems in non-collinear configurations}
\label{biGauge}

Reduced charts for $n$-body Hamiltonian systems in the non-collinear configurations region can be found by considering the Euclidean action of $SE(3)=\Rbb^3 \ltimes SO(3)$ on configuration space. It is because one considers the $SE(3)$ action on configuration space that the configurations must be non-collinear in order to avoid coordinate singularities when the action is not free. This is a different issue from the stratification of the reduced state space. Note that generally the configuration and state space isotropy subgroups for a given Lie group $G$ action are not the same, instead we have that $G_z \subset G_q$ for $z = (q,p)$. The gauge theoretic approach to cotangent bundle reduction is nicely reviewed by Littlejohn and Reinsch \cite{Littlejohn1997}. They however do not consider the final step required to reduce the rotational symmetry and fix the angular momentum. This is achieved by introducing Serret-Andoyer\footnote{Often also referred to as Andoyer or Deprit coodinates. A nice account of their history is given by Deprit and Elipe \cite{Deprit1993}.} coordinates, as explained by Deprit \cite{Deprit1967b} (see also \cite{Deprit1993,Ciftci2012}). These introduce inevitable coordinate singularities on the angular momentum sphere, which is probably why Littlejohn and Reinsch avoid them.
We shall briefly review the gauge theoretic approach for general $n$-body systems, and introduce our notation. 
A specific choice of charts for $n$-body system representing bimolecular reactions is given in Section \ref{spatialBimol}.

Consider a translation-reduced, rotation invariant $n$-body systems restricted to the non-collinear subset (i.e.~the trivial configuration isotropy-type submanifold) of 
configuration space $Q_{Id} \subset Q \cong \Rbb^{3(n-1)}$ and written in the Lagrangian formalism
\[L (R, \dot{R}) = \frac{1}{2} \sum_{i=1}^{n-1} m_i \vert \dot{R}_i \vert^2 - U(R),\]
where $R = (R_1, \cdots , R_{n-1})$ are some choice of Jacobi vectors, and $m_i$ the reduced masses.
The Jacobi vectors are not normalised or mass-weighted. We believe that the mass parameters are best dealt with by non-dimensionalising the system.
The potential $U$ is assumed to be invariant under the action of $SO(3)$.

Pass from the inertial frame $\lbrace X_1 ,X_2 ,X_3 \rbrace$ to a convenient rotating frame $\lbrace x_1 ,x_2 ,x_3 \rbrace$, which will depend on the problem at hand, and write
$ R_i = g \left( \psi \right) \cdot r_i,$ for $ i = 1, \cdots, n-1,$
where $g \in SO(3)$ is the rotation parametrised by the Euler angles $\psi = (\psi_1, \psi_2, \psi_3)$, and $r_i$ are the Jacobi vectors in the rotating frame. 
The rotating Jacobi vectors can be expressed in terms of $3n-6$ \emph{internal} coordinates $q$ for $Q_{Id}/SO(3)$ by specifying $r_i (q)$, which is called the \emph{gauge} in the physics literature \cite{Littlejohn1997}. 
We are effectively considering a fibre bundle $\pi_{Id}: Q_{Id} \rightarrow Q_{Id}/SO(3)$, and $q$ are coordinates for the base space. Then, $\sigma (q) = g \left( \psi \right) \cdot r_i (q)$ is a section, and the Euler angles $\psi = (\psi_1, \psi_2, \psi_3)$ are coordinates for the fibre, diffeomorphic to $SO(3)$.

In the new coordinates, the kinetic energy is
\[ 2 E_k = \sum_{i,j=1}^{3n-6} \dot{q}_i \tilde K_{ij} (q) \dot{q}_j + 2 \sum_{i,j=1}^{3} \sum_{k=1}^{3n-6} \omega_i I_{ij} (q) A_{kj} (q) \dot{q}_k + \sum_{i,j=1}^{3} \omega_i I_{ij} (q) \omega_j , \]
where $\omega$ is the angular velocity, that is the vector corresponding to the skew-symmetric matrix $\Omega (\psi) = g^T \left( \psi \right) \dot{g} \left( \psi \right)$, for which $\omega \times r = \Omega r$, for any 3-vector $r$. We are therefore considering an \emph{anholonomic} frame (or \emph{vielbein}) $(\dot q, \omega)$ for the tangent space at $(q,\psi)$, with $\omega = \Psi (\psi) \dot \psi$ \cite[Appendix C]{Littlejohn1997}.
The pseudo-metric $\tilde K (q)$ satisfies
 \[\tilde K_{ij} (q) = \sum_{k=1}^{n-1} m_k \frac{\del r_{k}(q)}{\del q_i} \cdot \frac{\del r_{k}(q)}{\del q_j}.\]
This is the restriction of the Euclidean metric on the (translation-reduced) configuration space $Q_{Id}$ to the section $\sigma(Q_{Id}/SO(3))$, and hence a ``pseudo-metric'' on the internal space $Q_{Id} /SO(3)$. It is of no importance in gauge theoretic terms, but nonetheless features prominently in the molecular literature.
The moment of inertia tensor $I (q)$ is given by
\[
I (q) = \sum_{k=1}^{n-1} m_k ( r_k (q)  \cdot r_k (q) I_d - r_k (q) \otimes r_k (q) ),
\]
or
\[I_{ij} (q) = \sum_{k=1}^{n-1} m_k \left( \sum_{s=1}^{3} r_{ks}(q)^2 \delta_{ij} - r_{ki}(q) r_{kj}(q) \right),\]
where $\otimes$ is the tensor, or outer, product for which 
$r_{ k} \otimes r_{ k} = r_{ k} r_{ k}^T$,
and the gauge potential $A (q)$ associated with the Coriolis effect, which is caused by the coupling term, is
\[A (q) = I^{-1}(q) a (q), \]
where $a (q) = (a_1(q), \cdots, a_{3n-6}(q) )$ and 
\[ a_i(q)  = \sum_{k=1}^{n-1} r_k(q) \times \frac{\del r_k(q)}{\del q_i}.\]
Equivalently
\[A_{ij} (q) = \sum_{k=1}^{n-1} \sum_{s,t,u=1}^{3} I^{js}(q) \epsilon_{stu} r_{kt}(q) \frac{\del r_{ku}(q)}{\del q_i},\]
where $I^{ks}(q)$ are components of $I^{-1}(q)$, and $\epsilon_{ijk}$ the Levi-Civita symbols\footnote{Recall, the Levi-Civita symbol $\epsilon_{ijk}$ is $1$ if $(i, j, k)$ is an even permutation of $(1,2,3)$, $-1$ if it is an odd permutation, and $0$ if any index is repeated.}.

The kinetic energy is gauge invariant, i.e.~independent of the choice of internal coordinates, but the individual terms are not (see \cite[Section IV.A]{Littlejohn1997}). It is therefore rewritten, in a gauge-invariant form, as
\[ 2E_k = \sum_{i,j=1}^{3n-6} \dot{q}_i K_{ij} (q) \dot{q}_j +
\sum_{i,j=1}^{3} \sum_{k=1}^{3n-6} \left( \omega_i + A_{ki} (q) \dot{q}_k \right) I_{ij} (q) \left( \omega_j + A_{kj} (q) \dot{q}_k \right) ,\]
where the metric $K(q) = \tilde K(q) - A^T(q) I(q) A(q)$ is an actual (Riemannian) metric on the internal space, obtained by projecting the metric on configuration space $Q_{Id}$ down to the internal space. It is therefore positive definite, but non-Euclidian due to the nature of the space \cite[Section IV.C]{Littlejohn1997}.

Finally, pass to the Hamiltonian formalism. The momenta are found (via the fibre derivative of the Lagrangian) to be
\begin{align*}
l_i &:= \frac{\del L (q, \psi, \dot q, \omega)}{\del \omega_i} = \sum_{j=1}^{3} \sum_{k=1}^{3n-6} I_{ij} (q) \left( \omega_j + A_{kj} (q) \dot{q}_k \right), \\
p_i &:= \frac{\del L (q, \psi, \dot q, \omega)}{\del \dot{q}_i} = 
\sum_{j=1}^{3n-6} K_{ij} (q) \dot{q}_j + \sum_{j=1}^{3} A_{ij} (q) l_j,
\end{align*}
where $l$ is the angular momentum in the rotating frame, i.e.~$l = g^T (\psi) \cdot L$.
The Hamiltonian is then the Legendre transform of the Lagrangian, namely
\begin{align*}
H ( q, \hat \psi, p, l ) &= \frac{1}{2} \sum_{i,j=1}^{3n-6} \sum_{k=1}^{3} \left( p_i - A_{ik} (q) l_k \right) K^{ij} (q) \left( p_j - A_{jk} (q) l_k \right) \\
&+ \frac{1}{2} \sum_{i,j=1}^{3} l_i I^{ij} (q) l_j + U \left( q \right),
\end{align*}
where the potential is a function of the internal coordinates only, due to the assumption of rotational invariance, and the Euler angles are ignorable.
The symplectic form is
\[\omega = \sum_{i=1}^{3n-6} \rmd q_i \wedge \rmd p_i + \sum_{i,j=1}^{3} \Psi_{ji} (\psi) \rmd \psi_i \wedge \rmd l_j + \frac{1}{2} \sum_{i,j,k,u,v=1}^{3} \epsilon_{ijk} l_i \Psi_{ju} (\psi) \Psi_{kv} (\psi) \rmd \psi_u \wedge \rmd \psi_v.
\]
Alternatively, the Poisson bracket for two smooth functions $F$, $G$ is
\[ \lbrace F,G \rbrace = \left( \del_{q_i} F \del_{p_i} G -  \del_{p_i} F \del_{q_i} G \right) + \Psi^{ji} \left( \del_{\psi_i} F \del_{l_j} G -  \del_{l_j} F \del_{\psi_i} G \right) - \epsilon_{ijk} l_i \del_{l_j} F \del_{l_k} G.\]
Littlejohn and Reinsch derive this in \cite[Section IV.D]{Littlejohn1997}.

The momenta $p$ are gauge dependent because of Coriolis term $A^T (q) l$. Passing to gauge-independent non-canonical momenta\footnote{Littlejohn and Reinsch call these ``covariant shape velocities'' and denote them $v$. We shall use the same notation, hoping that it will not lead to any confusion, even though it gives $v_i = K_{ij} (q) \dot{q}_j$.}, $v_i = p_i - A_{ij} (q) l_j$, simplifies the Hamiltonian to
\[H ( q, \hat \psi, v, l ) = \frac{1}{2} \sum_{i,j=1}^{3n-6} \sum_{k=1}^{3} v_i K^{ij} (q) v_j + V(q,l), \quad V(q,l) = 
\frac{1}{2} \sum_{i,j=1}^{3} l_i I^{ij} (q) l_j + U \left( q \right), \]
where $V$ is the effective potential combining the centrifugal term and the potential, 
and
\begin{align*}
\omega &= \sum_{i=1}^{3n-6} \rmd q_i \wedge \rmd v_i 
+ \sum_{i=1}^{3n-6} \sum_{j=1}^{3} A_{ij} (q) \rmd q_i \wedge \rmd l_j \\
&+ \frac{1}{2} \sum_{i,k=1}^{3n-6} \sum_{j=1}^{3} l_j  \left( B_{kij} (q) + \epsilon_{juv} A_{ku} (q) A_{iv} (q) \right) \rmd q_i \wedge \rmd q_k \\
&+ \sum_{i,j=1}^{3} \Psi_{ji} (\psi) \rmd \psi_i \wedge \rmd l_j + \frac{1}{2} \sum_{i,j,k,u,v=1}^{3} \epsilon_{ijk} l_i \Psi_{ju} (\psi) \Psi_{kv} (\psi) \rmd \psi_u \wedge \rmd \psi_v,
\end{align*}
where we have introduced the \emph{Coriolis tensor}
\[B_{ijk} (q) = \del_{q_i} A_{jk} (q) - \del_{q_j} A_{ik} (q) - \epsilon_{kst} A_{is} (q) A_{jt} (q) , \]
which is a curvature form on fibre bundle (see \cite[Section III.G]{Littlejohn1997}), and simplifies the equations of motion.
Effectively, this transformation moves the Coriolis effect from the Hamiltonian to the symplectic form, in the second and third terms. It is similar to using non-canonical coordinates for a charged particle in a magnetic field, such that the effect of the Lorentz force comes from the symplectic form, see e.g.~\cite[Section 2.10]{Marsden1992}.
The molecular literature usually does not pass to the gauge-invariant form of the kinetic energy, see discussion in \cite[Section IV.F]{Littlejohn1997}.

\begin{figure}
\begin{center}
	\def\JPicScale{1.2}
	\input{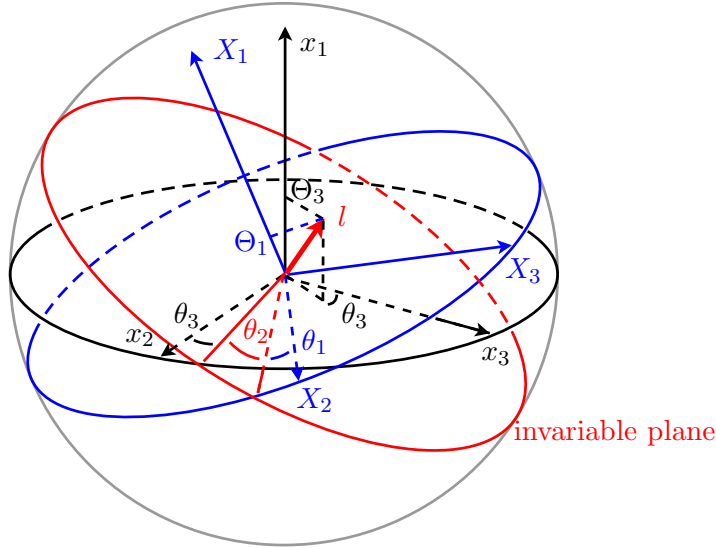}
\end{center} 
\caption{Transformation to Serret-Andoyer coordinates. $\lbrace X_1, X_2, X_3 \rbrace$ is the lab frame, $\lbrace x_1, x_2, x_3 \rbrace$ the chosen rotating frame and $l$ the angular momentum vector. $\Theta_2 = \vert l \vert$.}
\label{bimolAndo}
\end{figure}

By introducing the rotating frame, we have split the coordinates into internal coordinates $q$, (ignorable) rotations $\psi$ and their momenta, but we have not actually reduced the system. Since $(\psi, l)$ are non-canonical, the fact that $\psi$ are ignorable doesn't lead to $l$ being constant.
We can however pass from the non-canonical $(\psi, l)$ to canonical Serret-Andoyer coordinates $(\theta, \Theta)$ which consist of the total angular momentum $\vert l \vert$ plus two projections of $l$, which we are free to choose, and three angles. 
The choice of projection onto the $x_1$ and $X_1$-axis is shown in Figure \ref{bimolAndo}.

We immediately note that
\[l = l(\theta_3, \Theta_2, \Theta_3) = ( \Theta_3, \sqrt{ \Theta^2_2 -  \Theta_3^2} \sin \theta_3, \sqrt{ \Theta_2^2 -  \Theta_3^2} \cos \theta_3),\]
which we need to transform the Hamiltonian function, whereas the relations between the new angles $\theta$ and the non-canonical angular momentum coordinates is less straightforward and depends on the original choice of Euler angles. These are of no use to us here, but can be found in \cite{Deprit1967b} and \cite{Deprit1993}, where $(\theta, \Theta)$ are shown to be canonical coordinates.
The Hamiltonian in these new coordinates is
\[H ( q,\hat \theta_1 , \hat \theta_2 , \theta_3 , v, \hat \Theta_1, \Theta_2, \Theta_3 ) = \frac{1}{2} \sum_{i,j=1}^{3n-6} \sum_{k=1}^{3} v_i K^{ij} (q) v_j 
+ V(q,\theta_3, \Theta_2, \Theta_3). \]
Therefore, the system is reduced by eliminating the ignorable degree of freedom $(\theta_1, \Theta_1)$, fixing $\Theta_2 = \lambda$, which is the constant absolute value of the angular momentum, and eliminating $\theta_2$.
The remaining angular momentum coordinates $(\theta_3, \Theta_3)$ are the canonical latitude and longitude on the angular momentum sphere $\Sbb^2_\lambda$, henceforth denoted $z_\lambda = (q_\lambda, p_\lambda)$, and there is a coordinate singularity at $p_\lambda = \lambda$.
The reduced 
Hamiltonian function is
\begin{align*}
H \left( q, q_\lambda, v,p_\lambda; \lambda \right) &= \frac{1}{2} \sum_{i,j=1}^{3n-6} \sum_{k=1}^{3} v_i K^{ij} (q) v_j +
V(q,z_\lambda; \lambda) , \\
 V(q,z_\lambda; \lambda) &= 
 \frac{1}{2} \sum_{i,j=1}^{3} l_i (z_\lambda; \lambda) I^{ij} (q) l_j(z_\lambda; \lambda) + U \left( q \right),
 \end{align*}
and
\begin{align*}
\omega &= \sum_{i=1}^{3n-6} \rmd q_i \wedge \rmd v_i 
+ \sum_{i=1}^{3n-6} \sum_{j=1}^{3} A_{ij} (q) \del_{z_{\lambda k}} l_j (z_\lambda; \lambda) \rmd q_i \wedge \rmd z_{\lambda k} \\
& + \frac{1}{2} \sum_{i,k=1}^{3n-6} \sum_{j=1}^{3} l_j (z_\lambda; \lambda) \left( B_{kij} (q) + \epsilon_{juv} A_{ku} (q) A_{iv} (q) \right) \rmd q_i \wedge \rmd q_k + \rmd q_\lambda \wedge \rmd p_\lambda.
\end{align*}

The choice of projection is equivalent to a choice of which axis to use as a longitude for $\Sbb^2_\mu$.
The transformation for other projections is equivalent. By considering e.g.~minor and major principal axes, we get two charts that cover the whole of $\Sbb^2_\mu$.


\section{Spatial atom-molecule scaling}

\label{spatialDetials}

This appendix contains the scaling of the moment of inertia tensor, the gauge potential and the reduced metric for the spatial atom-diatom molecule capture problem of Section \ref{spatialBimol}.

The intermolecular distance \DoF is scaled as $x  = \varepsilon_c^{-1} \tilde x$ and $p_x = \varepsilon_c \tilde  p_x$, whereas the intramolecular coordinates are shifted such that $b=0$ at the equilibrium, and then scaled as $b = \varepsilon_b \tilde b$ and $p_b = \varepsilon_b^{-1} \tilde p_b$.
Thus, the scaled, rotating frame Jacobi vectors are
\begin{align*}
r_{n_b} (q) &= \varepsilon_c^{-1} \rho_{n_b} (x) = \varepsilon_c^{-1} x (1,0,0) \\
r_{i} (q) &= g_b  (\beta ) \cdot \rho_{ i}(b) = g_b  (\beta ) \cdot ( \rho_{ i}^0 + \varepsilon_b \sum_{j=1}^{3n_b-6} \rho_{ i j}^1 b_j) + \order{\varepsilon_b^2}, \quad  i = 1, \cdots , n_b - 1,
\end{align*}
where 
$\rho_{i}^{0} \in \Rbb^3$ are equilibrium configuration vectors, $g_b (\beta) \in SO(3)/SO(2)$ determines the orientation of $B$ and the $3(n_b-1)(3n_b-6)$ constants $\rho_{\beta ijk}^{1}$ determine the intramolecular coordinates $b$ and shall be chosen along the lines of the Eckart \cite{Eckart1935} and Sayvetz \cite{Sayvetz1939} conventions for normal and anomalous molecules in the small vibration regime.

If we write the moment of inertia tensor as
$
I (q) =: I_c (q)  + I_\beta (q), 
$
then
\[
I_\beta = \sum_{k=1}^{n_b -1} ( r_{ k} \cdot r_{ k} I_d - r_{ k} \otimes r_{ k}) 
= G_b ( \sum_{k=1}^{n_b -1} ( \rho_{ k} \cdot \rho_{ k} I_d - \rho_{ k} \otimes \rho_{ k} ) ) G_b^T,
\]
where $ G_b \rho_{ k} = g_b \cdot \rho_{ k} $.
Thus, it scales to
\[ I(q) = \varepsilon_c^{-2} I_c (x) + G_b (\beta) I_b^0 G_b^T (\beta) + \order{\varepsilon_b^1} \]
where $I_c (x) = \varepsilon_c^{-2} m_1 x^2 \text{Diag} (0,1,1)$, and we choose $I_b^0 = \text{Diag} ( \mu_{b1}, \mu_{b2}, \mu_{b3} )$ with $\mu_{b1} > \mu_{b2} > \mu_{b3}$,
via $g_b (\beta)$ and the rotation of $B$ about $x_1$, i.e.~$g (\psi)$. 
The inverse moment of inertia matrix exists,
since $I$ is a real, symmetric and positive definite for $B$ non-collinear, 
and scales as
\[
I^{-1} 
\sim  \left( \begin{array}{ccc}
 \varepsilon^0 & \varepsilon_c^2 & \varepsilon_c^2 \\
\varepsilon_c^2 & \varepsilon_c^2 & \varepsilon_c^4 \\
\varepsilon_c^2 & \varepsilon_c^4 & \varepsilon_c^2 \\
\end{array} \right) + \cdots.
\]

For our choice of coordinates and gauge, the gauge potential $A(q)$ has
\begin{align*}
a_x (q) &= 0, \\
a_{bi} (q) &= G_b (\beta ) \sum_{k=1}^{n_b -1} \rho^0_k \times \rho^1_{ki} + \varepsilon_b G_b (\beta ) \sum_{k=1}^{n_b -1} \sum_{j=1}^{3n_b -6} (\rho^1_{kj} \times \rho^1_{ki} ) b_j
= a_{bi}^0 (\beta) + \varepsilon_b a_{bi}^1 (\beta, b),\\
a_{\beta i} (q) &= a_{\beta i}^0 (\beta) + \order{\varepsilon_b},
\end{align*}
and we ask that $a_{b i}^0 (\beta) = 0$ for all $i = 1, \cdots, 3 n_b - 6$, i.e. 
\[\sum_{k=1}^{n_b-1} \left( \rho_{ k}^{0} \times \rho_{ ki}^{1} \right) = 0, \quad \forall i = 1, \cdots, 3 n_b - 6.\] 
This is known as the Eckart condition, and
imposes $3(3 n_b - 6)$ conditions on $\rho_{ kij}^{1}$.
The gauge potential scales to
\[ A(q) \sim 
\left( \begin{array}{ccc}
0 & \varepsilon_c^0 & 0 \\
0 & \varepsilon_c^2 &0 \\
0 & \varepsilon_c^2 &0
\end{array} \right) + \cdots,
\]
where $A_{\beta 1}^0 (\beta)$ and $A^0_{\beta i} (x,\beta)$ for $i = 2,3$.

Finally, for the reduced metric $K(q)$, which is a real, symmetric, positive definite matrix,
we consider the pseudo-metric $\tilde{K} (q)$ and write
$ \tilde{K} (q) = \tilde{K}_c (q) + \tilde{K}_\beta (q), $
where
\[ \tilde K_c = 
\left( \begin{array}{ccc}
1 & 0 & 0 \\
0 & 0  & 0 \\
0 & 0  & 0
\end{array} \right),
\quad
\tilde K_\beta = 
\left( \begin{array}{ccc }
0 & 0 & 0 \\
0 & \tilde K_\beta & \tilde K_{\beta b}\\
0 & \tilde K_{\beta b}^T & \tilde K_b 
\end{array} \right),
\]
and
\begin{align*}
\tilde K_{\beta ij} (q) &= \sum_{k=1}^{n_b -1} \frac{\del G_b}{\del \beta_i} \rho^0_k \cdot \frac{\del G_b}{\del \beta_j} \rho^0_k + \order{\varepsilon_b} = \tilde K^0_{\beta ij} (\beta) + \order{\varepsilon_b}\\
\tilde K_{\beta b ij} (q) &= \sum_{k=1}^{n_b -1} \frac{\del G_b}{\del \beta_i} \rho^0_k \cdot G_b \rho^1_{kj} + \order{\varepsilon_b} = \tilde K^0_{\beta b ij} (\beta) + \order{\varepsilon_b}\\
\tilde K_{b ij} (q) &= \sum_{k=1}^{n_b -1} \rho^1_{ki} \cdot \rho^1_{kj} + ... = \tilde K^0_{b ij} + ...
\end{align*}
We ask that $\tilde K_{b ij}^0 = ( \bar U_{\beta i}^2 )^{-1} \delta_{ij}$ for all $i,j$. That is we choose Williamson normal form coordinates for the intramolecular degrees of freedom.
These are $(3n_b-5)(3n_b-6)/2$ conditions on $\rho^1_{kij}$.
Furthermore, we claim that due to Eckart condition $\tilde K_{\beta b ij}^0 (\beta) = 0$ for all $i,j$, i.e.
\[ \sum_{k=1}^{n_b -1} \frac{\del G_b}{\del \beta_i} \rho^0_k \cdot G_b \rho^1_{kj} = 0. \]
Let us consider the case with $i = 1$. The Euler angles $\beta$ can be chosen in a number of ways, and the rotation matrix written as
$ G_b (\beta) = G_1 (\beta_1 ) G_2( \beta_2)$,
where $G_i( \beta_i)$ is a rotation by $\beta_i$ about some axis $y_i$. 
Recall that the symmetry about $x_1$ has been reduced and $G_b (\beta) \in \Sbb^2$.
Thus 
\[ \del_{\beta_1} G_b (\beta) = \del_{\beta_1} G_1 (\beta_1 ) G_2( \beta_2) =  G_1 (\beta_1 ) \tilde G_1( \frac{\pi}{2}) G_2( \beta_2),\]
where $\tilde G_1( \frac{\pi}{2})$ is a rotation about $y_1$ by $\frac{\pi}{2}$ and simultaneously a contraction in the $y_1$ direction to zero.
This can be seen by considering planar rotation matrices.
Then
\[ \sum_{k=1}^{n_b -1} \frac{\del G_b}{\del \beta_1} (\beta) \rho^0_k \cdot G_b (\beta) \rho^1_{kj} = \sum_{k=1}^{n_b -1} \tilde G_1( \frac{\pi}{2}) G_2( \beta_2) \rho^0_k \cdot G_2( \beta_2) \rho^1_{kj} = \sum_{k=1}^{n_b -1} \tilde G_1( \frac{\pi}{2}) \tilde \rho^0_k \cdot \tilde \rho^1_{kj}, \]
where $\tilde \rho_{kj}^i = G_2( \beta_2) \rho^i_{kj}$ and 
\[ \sum_{k=1}^{n_b -1} \tilde \rho_k^0 \times \tilde \rho_{kj}^0 = G_2( \beta_2) \sum_{k=1}^{n_b -1} \rho_k^0 \times \rho_{kj}^0 = 0, \]
by the Eckart condition. Thus $\tilde K_{\beta b 1j}^0 (\beta) = 0$, and the same is true for $i=2$.
The gauge dependent term of $K(q)$ instead scales to
\[ A^T I A \sim 
\left( \begin{array}{ccc}
0 & 0 &0 \\
0 & \varepsilon_c^0 & 0\\
0 & 0 &0
\end{array} \right) + \cdots,
\]
so
\[
K(q) = \left( \begin{array}{ccc}
1 & 0 & 0 \\
0 & \tilde K_\beta^0 (\beta) + F_0 (\beta) + \varepsilon_c^2 F_2 (x,\beta) + \varepsilon_c^4 F_4 (x,\beta) & 0\\
0 & 0 & \tilde D_b^{-1} 
\end{array} \right) + \order{\varepsilon_b}
\]
and
\[
K^{-1}(q)
= \left( \begin{array}{ccc}
1 & 0 & 0 \\
0 & K_{\beta 0}^{-1} (\beta) - \varepsilon_c^2 J_2 (x,\beta) + \varepsilon_c^4 J_4 (x,\beta) & 0\\
0 & 0 & \tilde D_b 
\end{array} \right) + \order{\varepsilon_b},
\]
by inverting the matrix blockwise, 
and expanding inverse matrices in 
formal power series.



\bibliographystyle{alphaDoi}
{\small
\newcommand{\etalchar}[1]{$^{#1}$}

}


\end{document}